\documentclass[11pt,a4paper]{article}

\usepackage[a4paper,margin=1in]{geometry}
\usepackage{setspace}
\usepackage[T1]{fontenc}
\usepackage[utf8]{inputenc}
\usepackage{lmodern}
\usepackage{microtype}
\usepackage{hyphenat}

\usepackage{amsmath,amssymb,amsfonts}
\usepackage{mathtools}
\usepackage{bm}
\usepackage{physics}
\usepackage{mathrsfs}

\usepackage{amsthm}

\theoremstyle{definition}

\theoremstyle{remark}

\usepackage{indentfirst}
\usepackage{graphicx}
\usepackage{float}
\usepackage{caption}
\usepackage{subcaption}
\usepackage{booktabs}
\usepackage{array}
\usepackage{multirow}
\usepackage{longtable}
\usepackage{tabularx}
\usepackage{adjustbox}

\usepackage{xcolor}
\usepackage[
    colorlinks=true,
    linkcolor=blue,
    citecolor=blue,
    urlcolor=blue
]{hyperref}
\usepackage{doi}
\usepackage[nameinlink,capitalize]{cleveref}

\usepackage[numbers,sort&compress]{natbib}

\usepackage{titlesec}

\titleformat{\section}
  {\normalfont\Large\bfseries}
  {\thesection.}
  {0.5em}
  {}

\titleformat{\subsection}
  {\normalfont\large\bfseries}
  {\thesubsection.}
  {0.5em}
  {}

\titleformat{\subsubsection}
  {\normalfont\normalsize\bfseries}
  {\thesubsubsection.}
  {0.5em}
  {}

\numberwithin{equation}{section}

\title{
    Fermionic Particle Creation in Curved Spacetime: Physical Bloch-Vector Dynamics and Exact de Sitter Scaling
}

\author{
    Samak Boonpan\\
    Department of Science and Technology,\\ Nampong Pattanasuksa Ratchamangklapisek School,\\
    Khon Kaen 40310, Thailand\\
}

\date{}

\begin{document}

\maketitle

\begin{abstract}
We formulate fermionic particle creation in a time-dependent curved spacetime as the exact evolution of a physical Bloch vector. Each Dirac momentum mode reduces to a two-level system, and its instantaneous occupation is obtained by projecting the physical state onto the instantaneous energy basis. This construction is equivalent to the standard Bogoliubov description while retaining the complete finite-time state trajectory in a compact real geometry. It also makes the fermionic occupation bound manifest without attributing its microscopic origin to
geometry alone. For a massive Dirac field in de Sitter spacetime, the dynamics reduces to a universal function of the physical momentum in Hubble units and the mass-to-Hubble ratio. We identify a mass-controlled nonadiabatic scale and a singular small-mass boundary layer that resolves
the noncommuting massless and late-time limits. Independent Cartesian Bloch, angular Bloch, and Bogoliubov integrations verify the analytical results and recover the standard Gibbons--Hawking Fermi--Dirac plateau. 
\end{abstract}


\section{Introduction}
\label{sec:introduction}

Particle creation is one of the most direct consequences of quantum
field theory in a time-dependent spacetime. The expansion of the
universe mixes positive- and negative-frequency modes and can transform
an initial vacuum into a state containing particles
\cite{Parker1968,Parker1969,BirrellDavies,ParkerToms,Ford2021}.
This mechanism appears in inflation, reheating, black-hole physics, and
other nonequilibrium gravitational backgrounds.

For fermionic fields, particle creation has several distinctive
features. The occupation number of each quantum state is bounded by the
Pauli exclusion principle, and a massless Dirac field in a conformally
flat spacetime does not undergo gravitational particle creation.
Nonzero mass, background interactions, or sufficiently rapid evolution
can nevertheless mix the particle and antiparticle sectors. Such
processes have been studied in fermionic preheating, axion inflation,
and de Sitter backgrounds
\cite{Greene1999,Adshead2015,Domcke2018,Cotaescu2002,Stahl2016}.

The usual description is based on Bogoliubov transformations. This method gives an invariant asymptotic particle number when well-defined in- and out-vacua exist. At a finite time, however, particle number depends on the reference basis used to separate positive- and negative-frequency modes
\cite{Fulling1973,BirrellDavies,DabrowskiDunne2016}.
Adiabatic constructions provide a systematic basis choice and are also essential for the renormalization of fermionic observables \cite{ParkerFulling1974,Landete2014,Ferreiro2022}. Nevertheless, the relation between finite-time occupation, unitarity, and Pauli blocking is not always transparent in the conventional mode-function description.

A complementary description follows from the two-level structure of each fermionic momentum and helicity sector. Pair production in an external electric field has been formulated through an underlying SU(2) algebra \cite{Seger1995}. Lewis--Riesenfeld invariants have also been extended to time-dependent quadratic fermionic systems \cite{Lewis1969,Kim1999}, and have been applied to explicitly time-dependent Dirac Hamiltonians \cite{Khantoul2015}. More recently, a group-theoretic formulation represented fermion production as the precession of an associated three-dimensional vector
\cite{Min2019}. These studies establish that SU(2) geometry is a natural language for fermionic evolution. More broadly, dynamical invariants underpin invariant-based inverse engineering and shortcuts to adiabaticity
\cite{GueryOdelin2019}.

The aim of the present work is therefore not to introduce the SU(2)
description itself. Instead, we develop an exact invariant formulation
in which the physical instantaneous Hamiltonian direction, the
invariant vector, and the state Bloch vector are carefully
distinguished. This distinction leads directly to a geometric
finite-time particle number and makes the Pauli bound manifest
throughout the evolution. It also avoids identifying the invariant
vector with the quantum-state Bloch vector, which would give an
incorrect particle interpretation.

We apply the construction to a massive Dirac field in a spatially flat
de Sitter universe. The dynamics reduces to a universal dimensionless
problem controlled by the mass ratio $m/H$ and the physical-momentum
variable $x=-k\eta$. This scaling separates genuine physical behavior
from the choice of comoving momentum. It also permits a direct analysis
of the conformal massless limit, the heavy-mass regime, the
nonadiabatic transition scale, and the late-time thermal occupation
associated with de Sitter spacetime
\cite{GibbonsHawking1977,Cotaescu2002}.

Real-time fermionic evolution in expanding or curved backgrounds is
also becoming accessible through digital quantum simulation and
spacetime-resolved computational quantum field theory
\cite{Gong2025,Alkhateeb2026}. These developments make exact
finite-time benchmarks increasingly useful. In the present work, the
Cartesian invariant equations, the angular representation, and the
Bogoliubov system are solved independently and shown to agree to
numerical precision.

The main results are the following. We obtain an exact SU(2) invariant
for each fermionic mode, derive the instantaneous occupation directly
from relative Bloch-sphere geometry, identify the de Sitter scaling
structure, and obtain the small-mass boundary-layer profile. The exact
Whittaker solution provides an independent analytical check of the
late-time occupation. The numerical analysis further reveals the
transition, overshoot, and relaxation of the finite-time particle
number before it reaches the de Sitter plateau.

The paper is organized as follows. Section~\ref{sec:dirac_flrw} reduces the Dirac field to independent two-level systems. Sections~\ref{sec:invariant_su2}
and~\ref{sec:pauli_blocking} construct the exact invariant and derive the geometric particle number. Section~\ref{sec:desitter} first establishes the conformal massless
benchmark and then develops the exact de Sitter scaling, its
nonadiabatic structure, and the relevant analytical limits.Section~\ref{sec:numerical_validation} presents the numerical results and cross-method validation. The final sections discuss the physical interpretation and summarize the conclusions.

\section{Fermionic Modes and the Instantaneous Particle Basis}
\label{sec:dirac_flrw}

We begin by reviewing the standard mode decomposition of a free massive spin-$1/2$ field in a spatially flat cosmological background. The primary purpose of this section is to establish the basic notation and to demonstrate explicitly how the spatial symmetries of the background spacetime allow each fermionic momentum mode to decouple. This decoupling is a crucial step, as it reduces the infinite-dimensional field dynamics into a set of independent two-level systems, each characterized by an $\mathrm{SU}(2)$ algebraic structure.

To describe the expanding background, we consider the spatially flat Friedmann--Lemaître--Robertson--Walker (FLRW) metric. It is mathematically convenient to express this metric in terms of the conformal time $\eta$, which yields
\begin{equation}
    ds^2 = a^2(\eta)\left(d\eta^2 - d\mathbf{x}^2\right),
    \label{eq:flrw_metric}
\end{equation}
where $a(\eta)$ is the time-dependent scale factor. We adopt the standard mostly-minus metric signature $(+,-,-,-)$. The use of conformal time makes the FLRW metric conformally equivalent to flat Minkowski spacetime. This coordinate choice simplifies the formulation of the covariant Dirac equation and facilitates the subsequent analysis of the massless limit, where the field theory exhibits conformal invariance.

\subsection{Covariant Dirac Equation and Mode Decomposition}

The dynamics of a free massive Dirac field $\psi(x)$ are governed by the covariant Dirac equation \cite{BirrellDavies}
\begin{equation}
    \left( i \gamma^\mu \nabla_\mu - m \right) \psi(x) = 0.
    \label{eq:covariant_dirac}
\end{equation}
To couple spinors to a curved background, it is necessary to introduce a local inertial frame at each spacetime point. This is achieved using the tetrad field $e^\mu_a$, which relates the curved spacetime gamma matrices $\gamma^\mu$ to the constant flat Minkowski gamma matrices $\tilde{\gamma}^a$ via the relation $\gamma^\mu=e^\mu_a\tilde{\gamma}^a$. For the spatially flat FLRW metric in Eq.~\eqref{eq:flrw_metric}, a standard choice for the tetrad is $e^\mu_a=a^{-1}(\eta)\delta^\mu_a$. Consequently, the curved gamma matrices are simply $\gamma^\mu = a^{-1}(\eta) \tilde{\gamma}^\mu$.

The spinor covariant derivative is defined as $\nabla_\mu=\partial_\mu+\Gamma_\mu$, where $\Gamma_\mu$ is the spin connection. The spin connection components are computed from the tetrad and the Christoffel symbols. For the FLRW metric in conformal time, the nonvanishing spin-connection components are
$ \Gamma_0=0, \Gamma_i = \frac{1}{2}\mathcal{H}\,  \widetilde{\gamma}_i\widetilde{\gamma}_0, \mathcal{H}
\equiv \frac{a'}{a},$ where $\mathcal{H}$ is the conformal Hubble parameter. Substituting these into the Dirac operator yields
\begin{equation}
    i \gamma^\mu \nabla_\mu 
    = i a^{-1} \tilde{\gamma}^\mu (\partial_\mu + \Gamma_\mu) 
    = i a^{-1} \tilde{\gamma}^\mu \partial_\mu + i \frac{3}{2} a^{-1} \mathcal{H} \tilde{\gamma}^0.
\end{equation}

In a spatially flat FLRW background, this spin-connection contribution can be entirely factored out by performing a standard conformal rescaling of the spinor field,
\begin{equation}
    \psi(x)=a^{-3/2}(\eta)\chi(x).
\end{equation}
When the temporal derivative acts on the scale factor $a^{-3/2}(\eta)$, it produces a term $-i \frac{3}{2} a^{-1} \mathcal{H} \tilde{\gamma}^0 \chi(x)$ which exactly cancels the spin-connection term. Multiplying the entire equation by $a(\eta)$, the equation of motion for the auxiliary field $\chi(x)$ reduces to a flat-space Dirac equation with a time-dependent effective mass:
\begin{equation}
    \left( i \tilde{\gamma}^\mu \partial_\mu - m a(\eta) \right) \chi(x) = 0.
    \label{eq:rescaled_dirac}
\end{equation}
Thus, the effect of the cosmological expansion is entirely encoded in the time-dependent mass term $m a(\eta)$.

Because the spatial sections of the FLRW metric are flat and translationally invariant, the spatial dependence of the field can be separated using plane waves. Taking the Fourier transform $\chi(x) = \int d^3k \, e^{i\mathbf{k}\cdot\mathbf{x}} \chi_{\mathbf{k}}(\eta)$, the momentum-space equation becomes
\begin{equation}
    \left( i \tilde{\gamma}^0 \partial_\eta - \tilde{\bm{\gamma}} \cdot \mathbf{k} - m a(\eta) \right) \chi_{\mathbf{k}}(\eta) = 0.
\end{equation}
Multiplying from the left by $\tilde{\gamma}^0$ and using the standard Dirac matrices $\tilde{\gamma}^0 = \beta$ and $\tilde{\bm{\gamma}} = \beta \tilde{\bm{\alpha}}$, we cast the equation into a Schrödinger form:
\begin{equation}
    i \partial_\eta \chi_{\mathbf{k}}(\eta) = \left( \tilde{\bm{\alpha}} \cdot \mathbf{k} + m a(\eta) \beta \right) \chi_{\mathbf{k}}(\eta).
    \label{eq:schrodinger_dirac}
\end{equation}

To decouple this four-component matrix equation, we construct the helicity operator $h = (\bm{\Sigma} \cdot \mathbf{k})/k$, where $\bm{\Sigma} = \mathrm{diag}(\bm{\sigma}, \bm{\sigma})$. The operator commutes with the Hamiltonian in Eq.~\eqref{eq:schrodinger_dirac}. We can therefore expand the rescaled field $\chi(x)$ using constant helicity eigenspinors $U_{\mathbf{k},h}$ and $V_{-\mathbf{k},h}$ as
\begin{equation}
    \chi(\eta,\mathbf{x})
    =
    \int \frac{d^3 k}{(2\pi)^{3/2}}
    e^{i\mathbf{k}\cdot\mathbf{x}}
    \sum_{h=\pm 1}
    \left[
        \hat{a}_{\mathbf{k},h} U_{\mathbf{k},h}(\eta)
        +
        \hat{b}^{\dagger}_{-\mathbf{k},h} V_{-\mathbf{k},h}(\eta)
    \right],
    \label{eq:mode_expansion}
\end{equation}
where $\hat{a}_{\mathbf{k},h}$ and $\hat{b}^{\dagger}_{-\mathbf{k},h}$ are the particle annihilation and antiparticle creation operators, respectively. Their nonvanishing equal-time canonical anticommutation relations are
\begin{equation*}
    \{\hat{a}_{\mathbf{k},h},\hat{a}^{\dagger}_{\mathbf{q},h'}\}
    =
    \{\hat{b}_{\mathbf{k},h},\hat{b}^{\dagger}_{\mathbf{q},h'}\}
    =
    \delta_{hh'}\delta^{(3)}(\mathbf{k}-\mathbf{q}),
\end{equation*}
with all mixed anticommutators vanishing.

Projecting onto a helicity eigenspinor reduces the four-component equation to an independent two-dimensional sector for each pair $(\mathbf{k},h)$. Before fixing a representation convention, its Hamiltonian can be written as
\begin{equation*}
    \hat H_{k,h}=hk\sigma_3+ma(\eta)\sigma_1.
\end{equation*}
The two helicity blocks are unitarily equivalent because
\begin{equation*}
    \hat H_{k,-}=\sigma_1\hat H_{k,+}\sigma_1.
\end{equation*}
The corresponding states and energy projectors transform by the same constant unitary matrix, so both helicities give the same occupation. We therefore display the representative sector $h=+1$. All occupation numbers and bounds below are understood per momentum-helicity sector; summing over both helicities introduces the usual factor of two. For $k=|\mathbf{k}|$, the positive- and negative-energy mode amplitudes are collected into the two-component vector
\begin{equation}
    \Psi_k(\eta)
    =
    \begin{pmatrix}
        u_k(\eta) \\
        v_k(\eta)
    \end{pmatrix}.
\end{equation}
The temporal evolution of this mode vector is governed by the reduced system of coupled first-order differential equations:
\begin{equation}
    i\frac{d}{d\eta}\Psi_k(\eta)
    =
    \hat{H}_k(\eta)\Psi_k(\eta).
    \label{eq:schrodinger_like}
\end{equation}
With the helicity convention specified above, a constant unitary choice of the two-component basis gives
\begin{equation}
    \hat{H}_k(\eta)
    =
    k\sigma_3 + m a(\eta)\sigma_1
    \equiv
    \mathbf{B}_k(\eta)\cdot\bm{\sigma},
    \label{eq:hamiltonian_su2}
\end{equation}
where $\bm{\sigma}=(\sigma_1,\sigma_2,\sigma_3)$ are the Pauli matrices. This block-diagonalization is standard; it does not constitute a new quantization procedure. The effective driving vector and conformal-time frequency are
\begin{equation}
    \mathbf{B}_k(\eta)
    =
    \left(m a(\eta),0,k\right),
    \qquad
    \omega_k(\eta)
    =
    |\mathbf{B}_k(\eta)|
    =
    \sqrt{k^2+m^2a^2(\eta)}.
    \label{eq:B_vector_frequency}
\end{equation}
The physical energy measured with respect to cosmic time is instead
\begin{equation}
    E_k(\eta)
    =
    \frac{\omega_k(\eta)}{a(\eta)}
    =
    \sqrt{\frac{k^2}{a^2(\eta)}+m^2}.
    \label{eq:physical_energy}
\end{equation}
Equation~\eqref{eq:hamiltonian_su2} shows that the exact mode dynamics are equivalent to those of a driven two-level system. This two-dimensional space contains the positive- and negative-energy amplitudes of one momentum-helicity sector; it is not the full fermionic Fock space. The associated particle--antiparticle Fock-state construction will be restored explicitly in Sec.~\ref{subsec:pauli_saturation}. This mode reduction provides the Bloch-vector representation used below.

\subsection{Instantaneous Vacuum and Bogoliubov Coefficients}
\label{subsec:bogoliubov}

At any fixed conformal time $\eta$, the instantaneous positive- and negative-energy spinors are defined as the local eigenstates of the effective Hamiltonian $\hat{H}_k(\eta)$,
\begin{equation}
    \hat{H}_k(\eta)\Psi_k^{(\pm)}(\eta)
    =
    \pm \omega_k(\eta)\Psi_k^{(\pm)}(\eta).
    \label{eq:instantaneous_eigenstates}
\end{equation}
These eigenstates define the instantaneous zeroth-order Hamiltonian basis used throughout this work. At finite times in a nonstationary spacetime, the resulting particle interpretation is basis dependent. To express the basis explicitly, it is convenient to parameterize the Hamiltonian by an angle $\vartheta_k(\eta)$ defined through $\tan \vartheta_k(\eta)=ma(\eta)/k$. The normalized positive- and negative-energy eigenvectors are then
\begin{equation}
    \Psi_k^{(+)}(\eta) = \begin{pmatrix} \cos(\vartheta_k/2) \\ \sin(\vartheta_k/2) \end{pmatrix}, \qquad 
    \Psi_k^{(-)}(\eta) = \begin{pmatrix} -\sin(\vartheta_k/2) \\ \cos(\vartheta_k/2) \end{pmatrix}.
\end{equation}
They provide a local particle--antiparticle splitting associated with instantaneous diagonalization of the conformal-time Hamiltonian. A higher-order adiabatic prescription would define a different finite-time basis and, in general, a different intermediate-time occupation and mismatch angle. This convention dependence disappears from the usual asymptotic Bogoliubov number when well-defined in- and out-frequency sectors exist \cite{DabrowskiDunne2016}.

Because the scale factor $a(\eta)$ evolves continuously, this adiabatic basis rotates within the two-dimensional state space. Consequently, an exact solution $\Psi_k(\eta)$ of the Schrödinger-like equation \eqref{eq:schrodinger_like} does not remain entirely in a single instantaneous eigenstate. It is standard practice to expand the general solution using a time-dependent Bogoliubov transformation:
\begin{equation}
    \Psi_k(\eta)
    =
    \alpha_k(\eta)\Psi_k^{(+)}(\eta)
    e^{-i\int_{\eta_0}^{\eta}\omega_k(\eta')\,d\eta'}
    +
    \beta_k(\eta)\Psi_k^{(-)}(\eta)
    e^{+i\int_{\eta_0}^{\eta}\omega_k(\eta')\,d\eta'} .
    \label{eq:bogoliubov_expansion}
\end{equation}
The exponential factors explicitly account for the standard dynamical phases accumulated during the temporal evolution. The complex functions $\alpha_k(\eta)$ and $\beta_k(\eta)$ are the Bogoliubov coefficients, which parameterize the mixing between the positive- and negative-energy sectors. 

By substituting the expansion \eqref{eq:bogoliubov_expansion} back into the differential equation \eqref{eq:schrodinger_like} and projecting the result onto the adiabatic basis vectors, the explicit temporal derivatives of the instantaneous eigenstates yield a system of coupled first-order ordinary differential equations:
\begin{align}
    \alpha_k'(\eta)
    &=
    \frac{\vartheta_k'(\eta)}{2}\,
    \beta_k(\eta)
    \exp\left(2i\int_{\eta_0}^{\eta}\omega_k(\eta')\,d\eta'\right),
    \label{eq:alpha_dot}
    \\
    \beta_k'(\eta)
    &=
    -\frac{\vartheta_k'(\eta)}{2}\,
    \alpha_k(\eta)
    \exp\left(-2i\int_{\eta_0}^{\eta}\omega_k(\eta')\,d\eta'\right).
    \label{eq:beta_dot}
\end{align}
Here a prime denotes $d/d\eta$, and the nonadiabatic coupling is
\begin{equation*}
    \vartheta_k'(\eta)
    =
    \frac{mka'(\eta)}{k^2+m^2a^2(\eta)}.
\end{equation*}
The mixing vanishes when the instantaneous Hamiltonian direction is time independent.

For fermionic fields, the canonical equal-time anticommutation relations impose a strict unitarity constraint on these mixing coefficients:
\begin{equation}
    |\alpha_k(\eta)|^2+|\beta_k(\eta)|^2=1.
    \label{eq:unitarity_bound}
\end{equation}
This algebraic normalization ensures probability conservation and fundamentally distinguishes fermions from bosons (which follow $|\alpha|^2 - |\beta|^2 = 1$). 

We impose the initial condition that the field is prepared in the instantaneous positive-energy state at a finite time $\eta_0$. For the positive-frequency mode $U_{\mathbf{k},h}(\eta)$, this gives $\alpha_k(\eta_0)=1$ and $\beta_k(\eta_0)=0$. In the asymptotic construction used later, $\eta_0$ is taken sufficiently early that this state converges to the positive-frequency in-vacuum. As the background evolves, the coupling $\vartheta_k'$ generates a nonzero negative-energy amplitude. Relative to the chosen instantaneous basis, the finite-time occupation number is

\begin{equation}
    N_k^{\mathrm{inst}}(\eta)
    =
    |\beta_k(\eta)|^2 .
    \label{eq:particle_number_def}
\end{equation}

At the algebraic level, Eq.~\eqref{eq:unitarity_bound} gives $0\leq N_k^{\mathrm{inst}}\leq1$ for each momentum-helicity sector. The fundamental origin of this bound is the canonical fermionic anticommutation algebra. The Bloch-sphere construction below gives an exact geometric representation of the same constraint.

The quantity $N_k^{\mathrm{inst}}(\eta)$ is therefore a basis-specified mode occupation, not a unique local particle observable. Only in an appropriate asymptotic regime, where positive- and negative-frequency sectors become unambiguous, does it reduce to the usual prescription-independent in/out particle number.

The Bogoliubov equations provide the standard complex-amplitude representation of the mode dynamics. In the following section we derive the equivalent real Bloch-vector evolution. The purpose of this reformulation is geometric transparency; no general computational superiority is assumed.

\section{Exact Bloch-Vector Dynamics}
\label{sec:invariant_su2}

The reduced Dirac equation is an exact two-level problem,
\begin{equation}
    i\frac{d}{d\eta}|\Psi_k(\eta)\rangle
    =
    \hat H_k(\eta)|\Psi_k(\eta)\rangle,
    \qquad
    \hat H_k(\eta)
    =
    \mathbf B_k(\eta)\cdot\bm{\sigma}.
    \label{eq:state_schrodinger}
\end{equation}
We formulate its evolution directly in terms of the physical state
density matrix rather than introducing an independent auxiliary
amplitude. The real-vector representation of two-level dynamics is
standard \cite{Feynman1957}; related phase-space formulations of the
Dirac vacuum likewise exploit a closed real system for fermionic
bilinears \cite{BialynickiBirula1991}.

\subsection{Physical state and exact precession}

For a normalized pure state,
\begin{equation}
    \rho_k(\eta)
    =
    |\Psi_k(\eta)\rangle\langle\Psi_k(\eta)|
    =
    \frac{1}{2}
    \left[
        \mathbb I+\mathbf n_k(\eta)\cdot\bm{\sigma}
    \right],
    \label{eq:physical_density_matrix}
\end{equation}
where
\begin{equation}
    \mathbf n_k(\eta)
    =
    \langle\Psi_k(\eta)|\bm{\sigma}|\Psi_k(\eta)\rangle
\end{equation}
is the physical state Bloch vector. Purity and normalization imply
\begin{equation}
    |\mathbf n_k(\eta)|=1.
    \label{eq:state_bloch_unit_norm}
\end{equation}

The density matrix satisfies
\begin{equation}
    \frac{d\rho_k}{d\eta}
    =
    -i[\hat H_k,\rho_k].
    \label{eq:state_liouville}
\end{equation}
Using
\begin{equation}
    [\sigma_i,\sigma_j]
    =
    2i\epsilon_{ijk}\sigma_k,
\end{equation}
Eq.~\eqref{eq:state_liouville} gives
\begin{equation}
    \frac{d\mathbf n_k}{d\eta}
    =
    2\mathbf B_k(\eta)\times\mathbf n_k(\eta).
    \label{eq:larmor_precession}
\end{equation}
This real precession equation is exactly equivalent to the two-component Dirac equation. No adiabatic or perturbative approximation has been made.

\subsection{Relation to a dynamical invariant}

A normalized Hermitian dynamical invariant may be written as
\begin{equation}
    \hat I_k(\eta)
    =
    \mathbf R_k(\eta)\cdot\bm{\sigma},
    \qquad
    |\mathbf R_k|=1,
    \label{eq:invariant_ansatz}
\end{equation}
and satisfies
\begin{equation}
    \frac{d\hat I_k}{d\eta}
    +
    i[\hat H_k,\hat I_k]
    =
    0.
    \label{eq:liouville}
\end{equation}
Consequently,
\begin{equation}
    \frac{d\mathbf R_k}{d\eta}
    =
    2\mathbf B_k\times\mathbf R_k.
    \label{eq:invariant_vector_equation}
\end{equation}

Equations~\eqref{eq:larmor_precession} and
\eqref{eq:invariant_vector_equation} have the same form, but this fact
alone does not identify an arbitrary invariant with the physical
state. Their initial data must also agree. For every pure solution of
the Schrödinger equation, the particular operator
\begin{equation}
    \hat I_{\Psi,k}
    \equiv
    2\rho_k-\mathbb I
    =
    \mathbf n_k\cdot\bm{\sigma}
    \label{eq:state_generated_invariant}
\end{equation}
is itself a dynamical invariant. In the present construction, the initial state is the instantaneous positive-energy eigenstate,
\begin{equation}
    \mathbf n_k(\eta_0)
    =
    \widehat{\mathbf B}_k(\eta_0).
    \label{eq:state_initial_alignment}
\end{equation}
If the invariant is initialized by the same alignment condition,
\begin{equation}
    \mathbf R_k(\eta_0)
    =
    \widehat{\mathbf B}_k(\eta_0),
\end{equation}
then uniqueness of the common precession equation gives
\begin{equation}
    \mathbf R_k(\eta)
    =
    \mathbf n_k(\eta)
\end{equation}
at all times. Thus, under the aligned initialization used here, the
selected invariant vector is the Bloch vector of the physical state
rather than an independent dynamical object. Other invariant solutions
with different initial directions remain mathematically valid but do
not encode the prepared state. We therefore use $\mathbf n_k$ as the
primary variable below.

\subsection{Angular coordinates}
\label{subsec:bloch_coordinates}

For analytical visualization, the unit state vector may be parameterized as
\begin{equation}
    \mathbf n_k(\eta)
    =
    \begin{pmatrix}
        \sin\Theta_k(\eta)\cos\Phi_k(\eta)\\
        \sin\Theta_k(\eta)\sin\Phi_k(\eta)\\
        \cos\Theta_k(\eta)
    \end{pmatrix}.
    \label{eq:bloch_vector}
\end{equation}
The normalized Hamiltonian direction is
\begin{equation}
    \widehat{\mathbf B}_k(\eta)
    =
    \begin{pmatrix}
        \sin\vartheta_k(\eta)\\
        0\\
        \cos\vartheta_k(\eta)
    \end{pmatrix},
    \qquad
    \tan\vartheta_k(\eta)
    =
    \frac{ma(\eta)}{k}.
    \label{eq:hamiltonian_angle}
\end{equation}

We define the instantaneous mismatch angle by
\begin{equation}
    \cos\gamma_k(\eta)
    =
    \widehat{\mathbf B}_k(\eta)\cdot\mathbf n_k(\eta).
    \label{eq:gamma_def}
\end{equation}
Explicitly,
\begin{equation}
    \cos\gamma_k
    =
    \sin\vartheta_k\sin\Theta_k\cos\Phi_k
    +
    \cos\vartheta_k\cos\Theta_k.
    \label{eq:gamma_explicit}
\end{equation}

Substitution of Eq.~\eqref{eq:bloch_vector} into the precession equation gives
\begin{align}
    \frac{d\Theta_k}{d\eta}
    &=
    -2ma(\eta)\sin\Phi_k,
    \label{eq:theta_dot}\\
    \frac{d\Phi_k}{d\eta}
    &=
    2k
    -
    2ma(\eta)\cot\Theta_k\cos\Phi_k.
    \label{eq:phi_dot}
\end{align}
The angular chart is singular at $\Theta_k=0,\pi$, although the Cartesian equation~\eqref{eq:larmor_precession} remains regular there. For this reason, the Cartesian form will be used as the primary equation in numerical validation.

\section{Geometric Occupation and Particle--Antiparticle Entanglement}
\label{sec:pauli_blocking}

\subsection{Exact projection formula}
\label{subsec:geometric_derivation}

The instantaneous negative-energy projector is
\begin{equation}
    P_{k,-}^{\mathrm{inst}}(\eta)
    =
    \frac{1}{2}
    \left[
        \mathbb I
        -
        \widehat{\mathbf B}_k(\eta)\cdot\bm{\sigma}
    \right].
    \label{eq:negative_energy_projector}
\end{equation}
The occupation relative to the chosen instantaneous basis is
\begin{equation}
    N_k^{\mathrm{inst}}(\eta)
    =
    \operatorname{Tr}
    \left[
        P_{k,-}^{\mathrm{inst}}(\eta)\rho_k(\eta)
    \right].
    \label{eq:geometric_projection}
\end{equation}
Using Eqs.~\eqref{eq:physical_density_matrix} and \eqref{eq:negative_energy_projector}, together with
\begin{equation}
    \operatorname{Tr}(\sigma_i)=0,
    \qquad
    \operatorname{Tr}(\sigma_i\sigma_j)=2\delta_{ij},
\end{equation}
we obtain
\begin{equation}
    N_k^{\mathrm{inst}}(\eta)
    =
    \frac{1}{2}
    \left[
        1-
        \widehat{\mathbf B}_k(\eta)\cdot\mathbf n_k(\eta)
    \right].
    \label{eq:master_formula_final}
\end{equation}
Equivalently,
\begin{equation}
    N_k^{\mathrm{inst}}(\eta)
    =
    \sin^2
    \left[
        \frac{\gamma_k(\eta)}{2}
    \right].
    \label{eq:beta_gamma_identity}
\end{equation}

Equation~\eqref{eq:master_formula_final} is exact once the instantaneous Hamiltonian basis has been specified. It does not remove the finite-time ambiguity of the particle concept in a nonstationary spacetime. Evaluating the same projector with the expansion in Eq.~\eqref{eq:bogoliubov_expansion} gives $N_k^{\mathrm{inst}}=|\beta_k|^2$. Thus the projection formula and the Bogoliubov definition are identical within the same instantaneous basis; the former is a geometric representation of the latter, not a different particle prescription.

Since both $\widehat{\mathbf B}_k$ and $\mathbf n_k$ are unit vectors,
\begin{equation}
    0
    \leq
    N_k^{\mathrm{inst}}
    \leq
    1.
    \label{eq:pauli_bound_geometric}
\end{equation}
The logical origin of this bound is the canonical anticommutation algebra, which produces a two-level fermionic sector and hence the normalized pure-state space $\mathbb{C}P^1\simeq S^2$. The Bloch sphere gives an exact geometric representation of the bound; it is not an independent derivation of the Pauli principle.

\subsection{Particle--antiparticle reduced state}
\label{subsec:pauli_saturation}

The density matrix in Eq.~\eqref{eq:physical_density_matrix} is pure:
\begin{equation}
    \rho_k^2=\rho_k,
    \qquad
    S_{\mathrm{vN}}(\rho_k)=0.
    \label{eq:pure_bloch_entropy}
\end{equation}
A nonzero binary entropy arises only after restoring the two-mode Fock structure and tracing over one member of the particle--antiparticle pair. This particle--antiparticle interpretation follows the standard
two-mode construction used for Dirac fields in expanding backgrounds
\cite{Fuentes2010,Kanno2017}.

For an independent momentum-helicity pair, the even-parity sector is spanned by the instantaneous vacuum and the particle--antiparticle pair. The corresponding pure state may be written as
\begin{equation}
    |\Omega_k(\eta)\rangle
    =
    \sqrt{1-N_k^{\mathrm{inst}}(\eta)}
    |0_k^{(p)}0_{-k}^{(a)}\rangle
    +
    e^{i\delta_k(\eta)}
    \sqrt{N_k^{\mathrm{inst}}(\eta)}
    |1_k^{(p)}1_{-k}^{(a)}\rangle.
    \label{eq:two_mode_fock_state}
\end{equation}
Here $\delta_k$ contains the convention-dependent relative phase. The full two-mode state is normalized and pure.
Tracing over the antiparticle sector gives
\begin{align}
    \rho_k^{(p)}
    &=
    \operatorname{Tr}_{a}
    |\Omega_k\rangle\langle\Omega_k|
    \nonumber\\
    &=
    \left(1-N_k^{\mathrm{inst}}\right)
    |0_k^{(p)}\rangle\langle0_k^{(p)}|
    +
    N_k^{\mathrm{inst}}
    |1_k^{(p)}\rangle\langle1_k^{(p)}|.
    \label{eq:particle_reduced_density}
\end{align}
The particle--antiparticle entanglement entropy is therefore
\begin{equation}
    S_k^{(p:a)}
    =
    -
    N_k^{\mathrm{inst}}\ln N_k^{\mathrm{inst}}
    -
    \left(1-N_k^{\mathrm{inst}}\right)
    \ln\left(1-N_k^{\mathrm{inst}}\right).
    \label{eq:von_neumann_entropy}
\end{equation}
In geometric form,
\begin{align}
    S_k^{(p:a)}
    &=
    -
    \sin^2\left(\frac{\gamma_k}{2}\right)
    \ln\left[
        \sin^2\left(\frac{\gamma_k}{2}\right)
    \right]
    \nonumber\\
    &\quad
    -
    \cos^2\left(\frac{\gamma_k}{2}\right)
    \ln\left[
        \cos^2\left(\frac{\gamma_k}{2}\right)
    \right].
    \label{eq:geometric_entanglement_entropy}
\end{align}
With the convention $0\ln0=0$, the entropy obeys $0\leq S_k^{(p:a)}\leq\ln2$ and reaches its maximum at $N_k^{\mathrm{inst}}=1/2$. In particular, complete occupation $N_k^{\mathrm{inst}}=1$ saturates the Pauli bound but is again a product state and therefore has zero particle--antiparticle entanglement. This entropy is a reparameterization of the occupation probability, not an additional independent observable. At finite times it inherits the basis dependence of the instantaneous particle--antiparticle splitting. It should not be interpreted as thermodynamic entropy production or decoherence of the full unitary state.


\section{Analytical Limits and Exact de Sitter Structure}
\label{sec:desitter}

We now examine the analytical structure of the mode dynamics. We first
use the conformally invariant massless theory as an exact consistency
benchmark. We then specialize to the expanding de Sitter patch and
derive the dimensionless scaling law, the mass-controlled
nonadiabatic scale, and the relation between the strictly massless and
late-time limits.

\subsection{Conformal massless benchmark}
\label{sec:conformal}

The strictly massless theory provides an exact consistency check of the
finite-time construction. Setting $m=0$ in the rescaled Dirac equation
\eqref{eq:rescaled_dirac} removes the scale factor from the mode
dynamics. The field therefore obeys the Minkowski-space Dirac equation
in conformal coordinates, as required by conformal invariance of a
massless Dirac field in a spatially flat FLRW spacetime.

For each momentum-helicity sector, the effective Hamiltonian reduces to

\begin{equation}
    \hat H_k
    =
    k\sigma_3,
    \qquad
    \mathbf B_k
    =
    k\widehat{\mathbf z}.
    \label{eq:massless-hamiltonian}
\end{equation}

Both the magnitude and the direction of the Hamiltonian are independent
of time. If the mode is initially prepared in the instantaneous
positive-energy state,

\begin{equation}
    \mathbf n_k(\eta_0)
    =
    \widehat{\mathbf z},
    \label{eq:massless-initial-state}
\end{equation}

the exact Bloch equation becomes

\begin{equation}
    \mathbf n_k'
    =
    2k\widehat{\mathbf z}
    \times
    \mathbf n_k.
    \label{eq:massless-bloch-equation}
\end{equation}

Because the initial state is parallel to the precession axis, the
right-hand side vanishes. Uniqueness of the initial-value problem then
gives

\begin{equation}
    \mathbf n_k(\eta)
    =
    \widehat{\mathbf z}
    \label{eq:massless-bloch-solution}
\end{equation}

for all conformal times. The state remains aligned with the
instantaneous positive-energy direction, and the projection formula
therefore yields

\begin{equation}
    N_k^{\mathrm{inst}}(\eta)
    =
    \frac{1}{2}
    \left[
        1-
        \widehat{\mathbf B}_k
        \cdot
        \mathbf n_k
    \right]
    =
    0.
    \label{eq:strict-massless-occupation}
\end{equation}

The same conclusion follows independently from the Bogoliubov system.
The instantaneous Hamiltonian angle satisfies

\begin{equation}
    \tan\vartheta_k(\eta)
    =
    \frac{ma(\eta)}{k}.
\end{equation}

Consequently, the strictly massless theory has

\begin{equation}
    \vartheta_k(\eta)=0,
    \qquad
    \vartheta_k'(\eta)=0.
    \label{eq:massless-zero-coupling}
\end{equation}

The positive- and negative-energy sectors therefore decouple.
Equations~\eqref{eq:alpha_dot} and \eqref{eq:beta_dot}, with the
positive-frequency initial condition, give

\begin{equation}
    \alpha_k(\eta)=1,
    \qquad
    \beta_k(\eta)=0.
    \label{eq:massless-bogoliubov-solution}
\end{equation}

Thus the Cartesian Bloch, angular, and Bogoliubov descriptions all
reproduce the standard conformal suppression of massless Dirac
particle creation. This result is stronger than the statement that an
asymptotic occupation vanishes: the finite-time occupation is
identically zero throughout the complete evolution. Expansion changes
the relation between comoving and physical momentum, but it cannot
rotate the Hamiltonian direction or mix the two energy sectors when
$m=0$.

This benchmark should not be interpreted as a new prediction. Its role
is to verify that the geometric projection preserves the conformal
behavior of the underlying Dirac theory. The strictly massless result
must nevertheless be distinguished from the limit obtained by evolving
a field with arbitrarily small but nonzero mass to the future de Sitter
boundary. The order of these two operations will be examined in
Sec.~\ref{sec:noncommuting-limits}.

\subsection{Dimensionless de Sitter reduction}
\label{subsec:de-sitter-reduction}

We now specialize to the expanding patch of de Sitter spacetime,

\begin{equation}
    a(\eta)
    =
    -\frac{1}{H\eta},
    \qquad
    \eta<0,
    \label{eq:de-sitter-scale-factor}
\end{equation}

where $H$ is the constant physical Hubble parameter. It is useful to
introduce the dimensionless variables

\begin{equation}
    x\equiv-k\eta>0,
    \qquad
    \nu\equiv\frac{m}{H}.
    \label{eq:de-sitter-dimensionless}
\end{equation}

The variable $x$ has a direct physical interpretation:

\begin{equation}
    x
    =
    \frac{k}{aH}
    =
    \frac{p_{\mathrm{phys}}}{H}.
    \label{eq:x-physical-meaning}
\end{equation}

It measures the physical momentum in Hubble units. Early conformal
times correspond to $x\gg1$, whereas the future boundary
$\eta\rightarrow0^{-}$ corresponds to $x\rightarrow0^{+}$.

In terms of $(x,\nu)$, the time-dependent mass becomes

\begin{equation}
    ma(\eta)
    =
    \frac{\nu k}{x}.
    \label{eq:scaled-mass}
\end{equation}

The conformal-time frequency is therefore

\begin{equation}
    \omega_k(\eta)
    =
    k
    \sqrt{
        1+\frac{\nu^2}{x^2}
    },
    \label{eq:scaled-conformal-frequency}
\end{equation}

while the physical energy is

\begin{equation}
    E_k(\eta)
    =
    \frac{\omega_k(\eta)}{a(\eta)}
    =
    H\sqrt{x^2+\nu^2}
    =
    \sqrt{
        p_{\mathrm{phys}}^2+m^2
    }.
    \label{eq:scaled-physical-energy}
\end{equation}

These relations show that the explicit dependence on the comoving
wavenumber can be removed from the evolution equation. The complete
mode trajectory is controlled only by the physical-momentum variable
$x$ and the mass ratio $\nu$. We now derive this scaling directly from
the exact Bloch equation.

\subsection{Exact de Sitter scaling}
\label{subsec:de-sitter-scaling}

The Hamiltonian vector can be factorized as

\begin{equation}
    \mathbf B_k(\eta)
    =
    k\mathbf b_\nu(x),
\end{equation}

where

\begin{equation}
    \mathbf b_\nu(x)
    =
    \left(
        \frac{\nu}{x},0,1
    \right).
    \label{eq:dimensionless-driving-vector}
\end{equation}

Since

\begin{equation}
    \frac{d}{d\eta}
    =
    -k\frac{d}{dx},
\end{equation}

the exact Bloch equation becomes

\begin{equation}
    \frac{d\mathbf n_k}{dx}
    =
    -2\mathbf b_\nu(x)
    \times
    \mathbf n_k(x).
    \label{eq:scaled-bloch-equation}
\end{equation}

The comoving momentum $k$ has disappeared. In Cartesian
components,

\begin{equation}
    \frac{dn_1}{dx}
    =
    2n_2,
    \label{eq:scaled-n1}
\end{equation}

\begin{equation}
    \frac{dn_2}{dx}
    =
    -2n_1
    +
    \frac{2\nu}{x}n_3,
    \label{eq:scaled-n2}
\end{equation}

\begin{equation}
    \frac{dn_3}{dx}
    =
    -\frac{2\nu}{x}n_2.
    \label{eq:scaled-n3}
\end{equation}

The normalized Hamiltonian direction is

\begin{equation}
    \widehat{\mathbf B}_\nu(x)
    =
    \frac{
        \left(
            \nu/x,0,1
        \right)
    }{
        \sqrt{1+\nu^2/x^2}
    }.
    \label{eq:scaled-hamiltonian-direction}
\end{equation}

At an initial value $x_0$, the instantaneous-vacuum condition is

\begin{equation}
    \mathbf n_\nu(x_0)
    =
    \widehat{\mathbf B}_\nu(x_0).
    \label{eq:scaled-initial-state}
\end{equation}

Both the differential equation and the initial condition depend
only on $x$, $x_0$, and $\nu$. In the asymptotic in-vacuum
limit $x_0\rightarrow\infty$, the solution is therefore a
universal trajectory $\mathbf n_\nu(x)$.

The instantaneous occupation has the scaling form

\begin{equation}
    N_k^{\mathrm{inst}}(\eta)
    =
    F_\nu(-k\eta),
    \label{eq:exact-scaling-law}
\end{equation}

where

\begin{equation}
    F_\nu(x)
    =
    \frac{1}{2}
    \left[
        1-
        \widehat{\mathbf B}_\nu(x)
        \cdot
        \mathbf n_\nu(x)
    \right].
    \label{eq:universal-occupation-function}
\end{equation}

Equation~\eqref{eq:exact-scaling-law} shows that calculations at
different $k$ but the same $x=-k\eta$ represent the same
trajectory. Different comoving momenta do not define independent
infrared, horizon, and ultraviolet dynamical regimes in exact
de Sitter spacetime.

The scaling law also distinguishes a finite-time momentum
distribution from an asymptotic spectrum. At a common finite
time $\eta_f$,

\begin{equation}
    N_k^{\mathrm{inst}}(\eta_f)
    =
    F_\nu(k|\eta_f|).
    \label{eq:finite-time-distribution}
\end{equation}

Large-$k$ modes then correspond to larger values of
$x_f=k|\eta_f|$ and may not yet have reached the late-time
region. Residual $k$-dependence at a common finite $\eta_f$ is
therefore a finite-time effect.

\subsection{Mass-controlled nonadiabaticity}
\label{subsec:nonadiabatic-scale}

The instantaneous Hamiltonian direction is parameterized by

\begin{equation}
    \tan\vartheta_\nu(x)
    =
    \frac{\nu}{x}.
    \label{eq:de-sitter-adiabatic-angle}
\end{equation}

Differentiation gives

\begin{equation}
    \frac{d\vartheta_\nu}{dx}
    =
    -\frac{\nu}{x^2+\nu^2}.
\end{equation}

Using $dx/d\eta=-k$, we obtain

\begin{equation}
    \vartheta_k'(\eta)
    =
    \frac{k\nu}{x^2+\nu^2}.
    \label{eq:de-sitter-angle-rate}
\end{equation}

A dimensionless measure of nonadiabaticity is

\begin{equation}
    \epsilon_{\mathrm{na}}(x)
    \equiv
    \frac{
        |\vartheta_k'|
    }{
        2\omega_k
    }.
\end{equation}

Substituting Eqs.~\eqref{eq:scaled-conformal-frequency} and
\eqref{eq:de-sitter-angle-rate} yields

\begin{equation}
    \epsilon_{\mathrm{na}}(x)
    =
    \frac{
        \nu x
    }{
        2(x^2+\nu^2)^{3/2}
    }.
    \label{eq:exact-nonadiabaticity}
\end{equation}

Its derivative is

\begin{equation}
    \frac{d\epsilon_{\mathrm{na}}}{dx}
    =
    \frac{
        \nu(\nu^2-2x^2)
    }{
        2(x^2+\nu^2)^{5/2}
    }.
\end{equation}

The unique maximum occurs at

\begin{equation}
    x_{\mathrm{peak}}
    =
    \frac{\nu}{\sqrt{2}}.
    \label{eq:nonadiabatic-peak}
\end{equation}

The maximum value is

\begin{equation}
    \epsilon_{\mathrm{na}}^{\mathrm{max}}
    =
    \frac{1}{3\sqrt{3}\,\nu}
    =
    \frac{H}{3\sqrt{3}\,m}.
    \label{eq:maximum-nonadiabaticity}
\end{equation}

Because $p_{\mathrm{phys}}=Hx$, the corresponding physical
momentum is

\begin{equation}
    p_{\mathrm{phys,peak}}
    =
    \frac{m}{\sqrt{2}}.
    \label{eq:mass-controlled-scale}
\end{equation}

The strongest rotation of the instantaneous particle basis is
therefore controlled by the fermion mass, not universally by the
de Sitter horizon. Horizon crossing occurs at $x=1$ and
coincides with the nonadiabatic peak only for $\nu=\sqrt{2}$.

For light fermions,

\begin{equation}
    \nu\ll1,
    \qquad
    x_{\mathrm{peak}}\ll1,
\end{equation}

so the strongest coupling occurs well after horizon crossing.
The condition for the maximum nonadiabaticity to exceed unity is

\begin{equation}
    \nu
    <
    \frac{1}{3\sqrt{3}}
    \simeq0.192.
    \label{eq:strong-nonadiabatic-regime}
\end{equation}

Equation~\eqref{eq:nonadiabatic-peak} locates the maximum of the
instantaneous coupling. The occupation number need not reach its
maximum at precisely the same point because its final value
depends on the coherent evolution over the complete trajectory.

\subsection{Noncommuting massless and late-time limits}
\label{sec:noncommuting-limits}

The strictly massless result in
Eq.~\eqref{eq:strict-massless-occupation} appears to differ from
the small-mass limit of the late-time de Sitter occupation. This
difference is caused by a nonuniform limit.

For every fixed $\nu>0$, the exact late-time result is

\begin{equation}
    \lim_{x\rightarrow0^+}
    F_\nu(x)
    =
    \frac{1}{e^{2\pi\nu}+1}.
    \label{eq:fixed-mass-late-limit}
\end{equation}

Taking $\nu\rightarrow0^+$ after the late-time limit gives

\begin{equation}
    \lim_{\nu\rightarrow0^+}
    \lim_{x\rightarrow0^+}
    F_\nu(x)
    =
    \frac{1}{2}.
    \label{eq:late-first-limit}
\end{equation}

By contrast, taking the massless limit first gives the
time-independent conformal theory,

\begin{equation}
    \lim_{x\rightarrow0^+}
    \lim_{\nu\rightarrow0^+}
    F_\nu(x)
    =
    0.
    \label{eq:mass-first-limit}
\end{equation}

Thus,

\begin{equation}
    \lim_{\nu\rightarrow0^+}
    \lim_{x\rightarrow0^+}
    F_\nu(x)
    \neq
    \lim_{x\rightarrow0^+}
    \lim_{\nu\rightarrow0^+}
    F_\nu(x).
    \label{eq:noncommuting-limits}
\end{equation}

The origin of this result is visible in the Hamiltonian
direction. At fixed $x>0$,

\begin{equation}
    \lim_{\nu\rightarrow0^+}
    \widehat{\mathbf B}_\nu(x)
    =
    \widehat{\mathbf z},
\end{equation}

whereas at fixed $\nu>0$,

\begin{equation}
    \lim_{x\rightarrow0^+}
    \widehat{\mathbf B}_\nu(x)
    =
    \widehat{\mathbf x}.
\end{equation}

The instantaneous basis used to define particles therefore has
different limits. The discontinuity is associated with the
particle basis, not with the unitarily evolving quantum state.

\subsubsection{Boundary layer at $x\sim\nu$}

The transition is resolved by introducing

\begin{equation}
    y
    \equiv
    \frac{x}{\nu}
    =
    \frac{p_{\mathrm{phys}}}{m}.
    \label{eq:boundary-layer-variable}
\end{equation}

The Hamiltonian direction becomes

\begin{equation}
    \widehat{\mathbf B}(y)
    =
    \frac{
        (1,0,y)
    }{
        \sqrt{1+y^2}
    }.
    \label{eq:boundary-layer-direction}
\end{equation}

It interpolates from $\widehat{\mathbf z}$ for $y\gg1$ to
$\widehat{\mathbf x}$ for $y\ll1$. In terms of $y$, the Bloch
equation is

\begin{equation}
    \frac{d\mathbf n_\nu}{dy}
    =
    -2\nu
    \left(
        \frac{1}{y},0,1
    \right)
    \times
    \mathbf n_\nu.
    \label{eq:boundary-layer-bloch}
\end{equation}

For fixed $y>0$ and $\nu\rightarrow0^+$, the state is frozen to
its incoming conformal direction:

\begin{equation}
    \mathbf n_\nu(\nu y)
    =
    \widehat{\mathbf z}
    +
    o(1).
    \label{eq:boundary-layer-state}
\end{equation}

The leading boundary-layer occupation is consequently

\begin{equation}
    F_{\mathrm{BL}}(y)
    =
    \frac{1}{2}
    \left[
        1-
        \widehat{\mathbf B}(y)
        \cdot
        \widehat{\mathbf z}
    \right],
\end{equation}

or

\begin{equation}
    F_{\mathrm{BL}}(y)
    =
    \frac{1}{2}
    \left(
        1-
        \frac{y}{\sqrt{1+y^2}}
    \right).
    \label{eq:boundary-layer-profile}
\end{equation}

For $y\gg1$,

\begin{equation}
    F_{\mathrm{BL}}(y)
    =
    \frac{1}{4y^2}
    +
    O(y^{-4}),
\end{equation}

while for $y\ll1$,

\begin{equation}
    F_{\mathrm{BL}}(y)
    =
    \frac{1}{2}
    -
    \frac{y}{2}
    +
    O(y^3).
\end{equation}

An arbitrarily small nonzero mass therefore produces a shrinking
transition layer at

\begin{equation}
    -k\eta
    =
    O\left(\frac{m}{H}\right),
\end{equation}

or equivalently at $p_{\mathrm{phys}}=O(m)$. Within this layer,
the instantaneous basis rotates by a finite angle while the
light-mass state cannot follow adiabatically.

\subsection{Late-time occupation and geometric angle}
\label{subsec:late-time-plateau}

The full spinor reconstruction in \ref{app:whittaker} gives, for fixed $\nu>0$,

\begin{equation}
    N_\nu^{\mathrm{out}}
    =
    \lim_{x\rightarrow0^+}
    F_\nu(x)
    =
    \frac{1}{e^{2\pi\nu}+1}.
    \label{eq:de-sitter-out-occupation}
\end{equation}

This is the Fermi--Dirac factor associated with the de Sitter
temperature

\begin{equation}
    T_{\mathrm{dS}}
    =
    \frac{H}{2\pi}.
\end{equation}

The result is a late-time physical-infrared limit at fixed
comoving momentum:

\begin{equation}
    \eta\rightarrow0^-,
    \qquad
    \frac{k}{a}\rightarrow0.
\end{equation}

It should not be described as an independent numerical
$k\rightarrow0$ spectrum.

The corresponding asymptotic geometric angle satisfies

\begin{equation}
    \sin^2
    \left(
        \frac{\gamma_{\mathrm{out}}}{2}
    \right)
    =
    \frac{1}{e^{2\pi\nu}+1}.
\end{equation}

Hence,

\begin{equation}
    \cos\gamma_{\mathrm{out}}
    =
    \tanh(\pi\nu),
    \label{eq:out-angle-cosh}
\end{equation}

and

\begin{equation}
    \gamma_{\mathrm{out}}
    =
    2\arctan
    \left(
        e^{-\pi\nu}
    \right).
    \label{eq:out-geometric-angle}
\end{equation}

For heavy fermions,

\begin{equation}
    \nu\gg1,
    \qquad
    N_\nu^{\mathrm{out}}
    \simeq
    e^{-2\pi\nu},
\end{equation}

and

\begin{equation}
    \gamma_{\mathrm{out}}
    \simeq
    2e^{-\pi\nu}.
\end{equation}

\subsection{Late-time particle--antiparticle entanglement}
\label{subsec:de-sitter-entanglement}

Using Eq.~\eqref{eq:von_neumann_entropy}, the reduced
particle--antiparticle entanglement entropy is

\begin{equation}
    S_\nu^{(p:a)}
    =
    -
    N_\nu^{\mathrm{out}}
    \ln N_\nu^{\mathrm{out}}
    -
    \left(
        1-N_\nu^{\mathrm{out}}
    \right)
    \ln
    \left(
        1-N_\nu^{\mathrm{out}}
    \right).
\end{equation}

Substitution of Eq.~\eqref{eq:de-sitter-out-occupation} gives

\begin{equation}
    S_\nu^{(p:a)}
    =
    \ln
    \left(
        1+e^{-2\pi\nu}
    \right)
    +
    \frac{
        2\pi\nu
    }{
        e^{2\pi\nu}+1
    }.
    \label{eq:de-sitter-entanglement}
\end{equation}

For heavy fermions,

\begin{equation}
    S_\nu^{(p:a)}
    \simeq
    \left(
        1+2\pi\nu
    \right)
    e^{-2\pi\nu}.
\end{equation}

If the late-time limit is taken before the small-mass limit,

\begin{equation}
    \lim_{\nu\rightarrow0^+}
    S_\nu^{(p:a)}
    =
    \ln2.
\end{equation}

The strictly massless theory instead has

\begin{equation}
    S_0^{(p:a)}(\eta)
    =
    0
\end{equation}

at every finite time. The entropy therefore inherits the
noncommuting limits:

\begin{equation}
    \lim_{\nu\rightarrow0^+}
    \lim_{x\rightarrow0^+}
    S_\nu^{(p:a)}(x)
    =
    \ln2,
\end{equation}

whereas

\begin{equation}
    \lim_{x\rightarrow0^+}
    \lim_{\nu\rightarrow0^+}
    S_\nu^{(p:a)}(x)
    =
    0.
    \label{eq:entropy-noncommuting-limits}
\end{equation}

The value $\ln2$ is not thermodynamic entropy production in the
strictly massless conformal theory. It is the reduced
particle--antiparticle entanglement obtained by first taking the
late-time limit of a nonzero-mass theory.


\section{Numerical Validation and Universal de Sitter Dynamics}
\label{sec:numerical_validation}

We now validate the numerical implementation against the analytical
results of the preceding section. The purpose is threefold. First, we
verify the exact de Sitter scaling in the variable
$x=-k\eta$. Second, we compare the Cartesian Bloch, angular Bloch, and
Bogoliubov formulations throughout the finite-time evolution. Third,
we examine the small-mass boundary layer associated with the noncommuting massless
and late-time limits.

\subsection{Numerical setup}
\label{subsec:numerical_setup}

We use the dimensionless variables $x=-k\eta>0$ and $\nu=m/H$
introduced in Eq.~\eqref{eq:de-sitter-dimensionless}. The exact de
Sitter equations are then independent of the comoving wavenumber $k$.
The Cartesian physical Bloch vector satisfies
\begin{equation}
    \frac{d\mathbf{n}}{dx}
    =
    -2\,\mathbf{b}_{\nu}(x)\times\mathbf{n},
    \qquad
    \mathbf{b}_{\nu}(x)
    =
    \left(
    \frac{\nu}{x},0,1
    \right).
    \label{eq:numerical_cartesian}
\end{equation}
At a finite initial point $x_0$, the state is initialized along the
instantaneous positive-energy direction,
\begin{equation}
    \mathbf{n}(x_0)
    =
    \widehat{\mathbf{b}}_{\nu}(x_0).
    \label{eq:numerical_initial_state}
\end{equation}
The instantaneous occupation is evaluated from
\begin{equation}
    N_k^{\mathrm{inst}}(x)
    =
    \frac{1}{2}
    \left[
    1-
    \widehat{\mathbf{b}}_{\nu}(x)\cdot\mathbf{n}(x)
    \right].
    \label{eq:numerical_occupation}
\end{equation}

The displayed trajectories were obtained with the DOP853 \cite{Hairer1993} adaptive
integrator using relative and absolute tolerances
\begin{equation}
    \mathrm{rtol}=10^{-11},
    \qquad
    \mathrm{atol}=10^{-13}.
    \label{eq:numerical_tolerances}
\end{equation}
The solution was returned at $2800$ points on a decreasing logarithmic
grid from $x_0=10^3$ to $x_f=10^{-8}$. These output points do not set
the internal step size, which was selected by the embedded error
estimator. No fixed step size or manual step-size cap was imposed in the
production runs. Neither the Bloch vector nor the
Bogoliubov amplitudes were renormalized after integration, and the
computed occupation was not clipped to the interval $[0,1]$.
Normalization and unitarity were instead monitored as independent
diagnostics.

For cross-validation, we also integrated the angular equations
\begin{align}
    \frac{d\Theta}{dx}
    &=
    \frac{2\nu}{x}\sin\Phi,
    \label{eq:numerical_theta}
    \\
    \frac{d\Phi}{dx}
    &=
    -2+
    \frac{2\nu}{x}
    \cot\Theta\cos\Phi,
    \label{eq:numerical_phi}
\end{align}
with
\begin{equation}
    \Theta(x_0)=\arctan\left(\frac{\nu}{x_0}\right),
    \qquad
    \Phi(x_0)=0.
\end{equation}
Because this angular chart becomes ill-conditioned near the poles, it
is used here as a cross-check rather than as the primary numerical
representation. The Cartesian equation \eqref{eq:numerical_cartesian}
is globally regular.

A third calculation was performed using the Bogoliubov amplitudes,
\begin{align}
    \frac{d\alpha}{dx}
    &=
    -C_\nu(x)\,
    \beta\,e^{2i\varphi},
    \\
    \frac{d\beta}{dx}
    &=
    C_\nu(x)\,
    \alpha\,e^{-2i\varphi},
    \\
    \frac{d\varphi}{dx}
    &=
    -\sqrt{1+\frac{\nu^2}{x^2}},
\end{align}
where
\begin{equation}
    C_\nu(x)
    =
    \frac{\nu}{2(x^2+\nu^2)},
    \qquad
    \alpha(x_0)=1,
    \qquad
    \beta(x_0)=0,
    \qquad
    \varphi(x_0)=0.
\end{equation}
In this representation, the occupation is
$N_k^{\mathrm{inst}}=|\beta|^2$.

\subsection{Cross-validation of the three formulations}
\label{subsec:cross_validation}

\begin{figure*}[t]
    \centering
    \includegraphics[width=\textwidth]{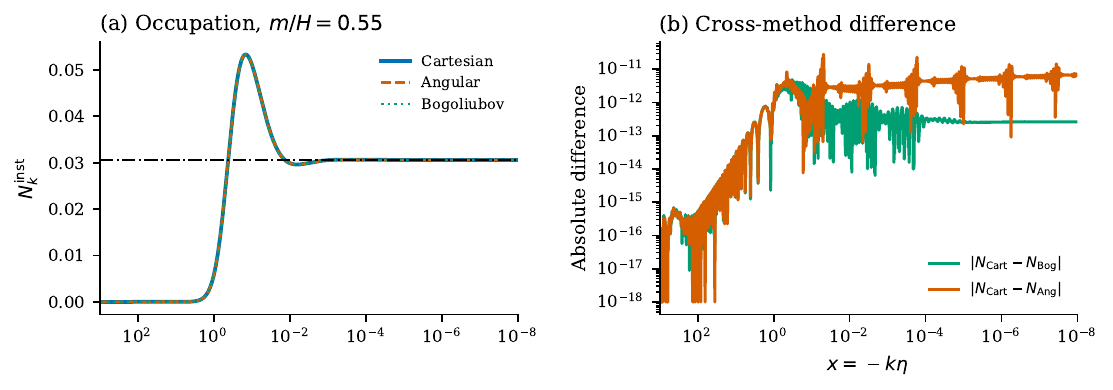}
    \caption{
    Independent numerical validation for $\nu=m/H=0.55$.
    (a) Instantaneous occupation obtained from the Cartesian Bloch
    equation, the angular Bloch equations, and the Bogoliubov
    amplitudes. The horizontal dash-dotted line is the exact late-time
    value $[e^{2\pi\nu}+1]^{-1}$.
    (b) Absolute differences between the three calculations.
    For this mass, the pairwise differences remain below
    $3\times10^{-11}$ over the full evolution.
    }
    \label{fig:method_validation}
\end{figure*}

Figure~\ref{fig:method_validation} compares three algebraically
equivalent but independently integrated representations for the
intermediate mass $\nu=0.55$. The Cartesian, angular, and Bogoliubov
curves are visually indistinguishable over the complete evolution. The
lower panel displays their absolute differences. This comparison tests
the implementation across different dynamical variables; agreement
with the analytical plateau provides the separate physical benchmark.

For the three masses used below, the largest differences are
\begin{equation}
    \max_x
    \left|
    N_{\mathrm{Cart}}-N_{\mathrm{Bog}}
    \right|
    \leq
    1.7\times10^{-10},
\end{equation}
and
\begin{equation}
    \max_x
    \left|
    N_{\mathrm{Cart}}-N_{\mathrm{Ang}}
    \right|
    \leq
    2.8\times10^{-11}.
\end{equation}
The Cartesian norm drift remains below
$9\times10^{-12}$, while the Bogoliubov unitarity error satisfies
\begin{equation}
    \max_x
    \left|
    |\alpha|^2+|\beta|^2-1
    \right|
    \leq
    1.8\times10^{-11}.
\end{equation}
The transient structures discussed below therefore occur in all three
representations and are not produced by numerical norm drift.

\subsection{Initial- and final-time convergence}
\label{subsec:numerical_convergence}

The finite numerical boundaries require separate convergence tests.
We define the relative error with respect to the exact late-time
plateau by
\begin{equation}
    \delta(x_0,x_f)
    =
    \frac{
    \left|
    N_k^{\mathrm{inst}}(x_f)
    -
    N_\infty(\nu)
    \right|
    }{
    N_\infty(\nu)
    },
    \qquad
    N_\infty(\nu)
    =
    \frac{1}{e^{2\pi\nu}+1}.
    \label{eq:relative_numerical_error}
\end{equation}
Table~\ref{tab:numerical_convergence} reports the dependence on both
the initial point $x_0$ and the stopping point $x_f$.

\begin{table*}[t]
\centering
\caption{
Convergence of the Cartesian Bloch integration. The entries are the
relative errors $\delta$ defined in
Eq.~\eqref{eq:relative_numerical_error}.
The upper block varies $x_0$ at fixed $x_f=10^{-8}$.
The lower block varies $x_f$ at fixed $x_0=4000$.
}
\label{tab:numerical_convergence}

\begin{tabular}{c|ccccc}
\hline\hline
$\nu$ &
$x_0=20$ &
$x_0=100$ &
$x_0=500$ &
$x_0=1000$ &
$x_0=4000$
\\
\hline
$0.05$ &
$5.68\times10^{-5}$ &
$2.51\times10^{-6}$ &
$1.53\times10^{-7}$ &
$8.36\times10^{-8}$ &
$5.81\times10^{-8}$
\\
$0.55$ &
$3.04\times10^{-3}$ &
$1.36\times10^{-4}$ &
$5.10\times10^{-6}$ &
$1.43\times10^{-6}$ &
$8.64\times10^{-8}$
\\
$1.70$ &
$4.24\times10^{-1}$ &
$1.57\times10^{-2}$ &
$5.84\times10^{-4}$ &
$1.65\times10^{-4}$ &
$1.12\times10^{-5}$
\\
\hline
\end{tabular}

\vspace{0.8em}

\begin{tabular}{c|cccc}
\hline
$\nu$ &
$x_f=10^{-2}$ &
$x_f=10^{-4}$ &
$x_f=10^{-6}$ &
$x_f=10^{-8}$
\\
\hline
$0.05$ &
$2.04\times10^{-1}$ &
$1.46\times10^{-3}$ &
$4.99\times10^{-6}$ &
$5.81\times10^{-8}$
\\
$0.55$ &
$2.35\times10^{-2}$ &
$5.18\times10^{-4}$ &
$6.14\times10^{-6}$ &
$8.64\times10^{-8}$
\\
$1.70$ &
$1.28\times10^{-1}$ &
$1.44\times10^{-3}$ &
$1.59\times10^{-6}$ &
$1.12\times10^{-5}$
\\
\hline\hline
\end{tabular}
\end{table*}

The heavy-mass case is the most sensitive to the finite initial
boundary. For $\nu=1.70$, starting at $x_0=20$ produces an error of
approximately $42\%$, whereas increasing the initial point to
$x_0=4000$ reduces the error to approximately $10^{-5}$. At the
smallest stopping points, the remaining error is controlled mainly by
the finite-$x_0$ initialization. Small non-monotonic changes in the
last digits therefore reflect residual finite-$x_0$ phase sensitivity,
rather than loss of numerical stability. This interpretation is
supported by the norm and cross-method diagnostics above.

We also tested convergence with respect to the solver controls for the
representative case $\nu=0.55$. Relative to a calculation with
$\mathrm{rtol}=2.5\times10^{-14}$ and
$\mathrm{atol}=2.5\times10^{-16}$, the maximum difference over the
complete trajectory decreases from $3.6\times10^{-9}$ at
$(\mathrm{rtol},\mathrm{atol})=(10^{-8},10^{-10})$ to
$4.3\times10^{-13}$ at $(10^{-12},10^{-14})$. Imposing the additional
step-size caps $\Delta x_{\max}=0.5$ and $0.25$ changes the production
trajectory obtained with the tolerances in
Eq.~\eqref{eq:numerical_tolerances} by at most
$4.1\times10^{-12}$. The unrestricted adaptive result is therefore
already stable at the accuracy required for the figures.

\subsection{Universal finite-time trajectories}
\label{subsec:universal_trajectories}

\begin{figure*}[t]
    \centering
    \includegraphics[width=\textwidth]{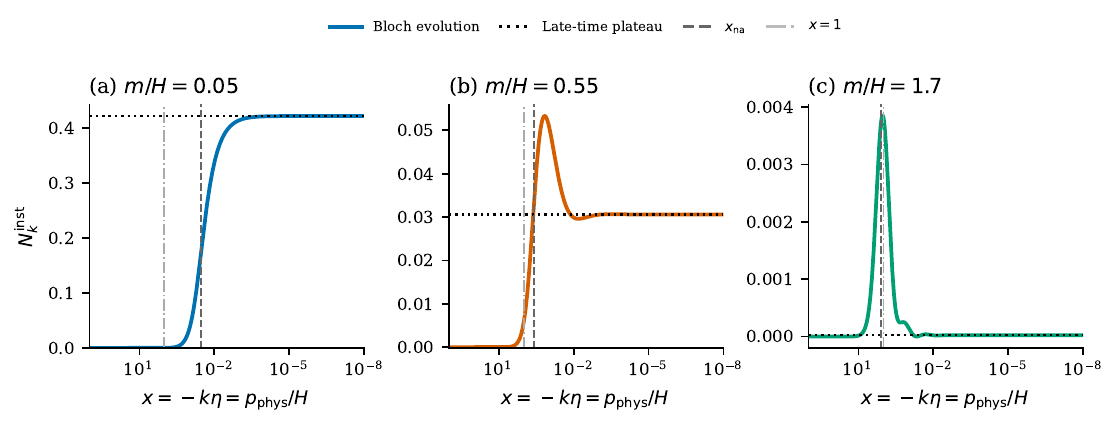}
    \caption{
    Universal finite-time occupation as a function of
    $x=-k\eta=p_{\mathrm{phys}}/H$.
    Panels (a)--(c) correspond to
    $\nu=0.05$, $0.55$, and $1.70$, respectively.
    The dotted line is the exact late-time plateau
    $N_\infty(\nu)=[e^{2\pi\nu}+1]^{-1}$.
    The dashed line marks the maximum of the non-adiabaticity parameter,
    $x_{\mathrm{na}}=\nu/\sqrt{2}$, giving
    $x_{\mathrm{na}}=0.0354$, $0.389$, and $1.20$.
    The light dash-dotted line at $x=1$ denotes the Hubble scale.
    Time evolution proceeds from left to right because $x$ decreases
    during expansion.
    }
    \label{fig:universal_trajectories}
\end{figure*}

The exact scaling law in Eq.~\eqref{eq:exact-scaling-law} implies that
comoving momenta do not define independent dynamical curves in exact de
Sitter space. At fixed $\nu$, changing $k$ only changes the mapping
between conformal time and $x=-k\eta$; it does not generate a new
function $F_\nu(x)$. Figure~\ref{fig:universal_trajectories} therefore
displays one universal trajectory for each value of $\nu=m/H$.

For the light mass $\nu=0.05$, the occupation grows smoothly toward the
exact late-time value $N_\infty\simeq0.42210$. The transition is
controlled by the mass-dependent scale $x\sim\nu$, rather than
universally by horizon crossing at $x=1$, consistently with the
analytical maximum $x_{\mathrm{na}}=\nu/\sqrt{2}$.

For the intermediate mass $\nu=0.55$, the occupation exhibits a
transient overshoot before approaching
$N_\infty\simeq0.030598$. The same overshoot appears in the Cartesian,
angular, and Bogoliubov calculations shown in
Fig.~\ref{fig:method_validation}. For the heavier mass $\nu=1.70$, the
finite-time response is concentrated near $x\sim1$ and reaches a
transient value of order $10^{-3}$, whereas its asymptotic occupation
is only $N_\infty\simeq2.30\times10^{-5}$. The finite-time occupation
relative to the chosen instantaneous basis can therefore be much
larger than its final asymptotic value.

These transient curves should be interpreted with care.
The quantity $N_k^{\mathrm{inst}}(x)$ is defined relative to the
instantaneous zeroth-order energy basis. Its detailed transient shape,
including the overshoot and coherent ringdown, is therefore
prescription-dependent. The late-time value has a more robust
asymptotic particle interpretation, whereas the intermediate-time
trajectory describes the evolution associated with the basis adopted
in this work. It should not be interpreted as a unique local particle
number in a nonstationary spacetime.

\subsection{Small-mass boundary layer}
\label{subsec:numerical_boundary_layer}

\begin{figure}[t]
    \centering
    \includegraphics[width=0.95\columnwidth]
    {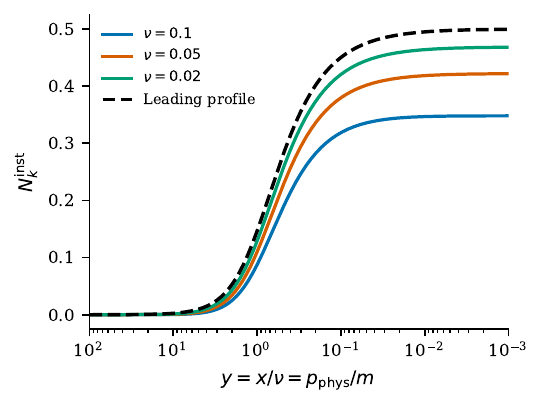}
    \caption{
    Approach to the small-mass boundary-layer profile.
    The numerical solutions are plotted against
    $y=x/\nu=p_{\mathrm{phys}}/m$ for
    $\nu=0.10$, $0.05$, and $0.02$.
    As $\nu$ decreases, the finite-mass curves approach the leading
    profile in Eq.~\eqref{eq:boundary-layer-profile}.
    Their late-time values remain below $1/2$ at finite $\nu$ and
    approach $1/2$ only as $\nu\rightarrow0^+$.
    }
    \label{fig:boundary_layer}
\end{figure}

The analytical small-mass limit is controlled by the stretched
variable $y=x/\nu=p_{\mathrm{phys}}/m$ defined in
Eq.~\eqref{eq:boundary-layer-variable}. Figure~\ref{fig:boundary_layer}
compares the leading profile in Eq.~\eqref{eq:boundary-layer-profile}
with numerical trajectories at decreasing positive masses.

The numerical sequence approaches the predicted profile over the
complete transition region $y=O(1)$, while the finite-mass plateaus
move toward $1/2$ as $\nu\to0^+$. This behavior does not contradict the
exact massless result $N_k^{\mathrm{inst}}=0$. The strict massless
problem has a fixed Hamiltonian direction and is not described by
setting $\nu=0$ in the stretched coordinate $y=x/\nu$. The numerical
collapse therefore confirms the noncommuting limits stated in
Eq.~\eqref{eq:noncommuting-limits} without repeating their analytical
derivation.

The numerical results thus confirm four analytical properties:
the exact reduction to $F_\nu(-k\eta)$, the mass-dependent
non-adiabatic scale $x_{\mathrm{na}}=\nu/\sqrt{2}$, the standard
Gibbons--Hawking late-time plateau, and the singular small-mass
boundary layer. At the same time, the independently integrated
Cartesian, angular, and Bogoliubov representations provide a direct reproducibility
check of the finite-time evolution.


\section{Discussion, Scope, and Applications}
\label{sec:discussion}

The present formulation does not modify the standard quantization of
the Dirac field or introduce a new definition of particle number. Its
purpose is to reorganize the same unitary mode dynamics into a compact
geometric representation. The Bogoliubov amplitudes, the physical
Bloch vector, and the instantaneous occupation describe the same
two-level state from different viewpoints. The numerical agreement
demonstrated in Section~\ref{sec:numerical_validation} confirms this
equivalence throughout the finite-time evolution, rather than only in
the asymptotic limit.

\subsection{Established ingredients and new results}

The reduction of a fermionic mode to an $\mathrm{SU}(2)$ system, its
Bogoliubov description, and the Bloch representation of two-level
dynamics are established ingredients. We do not claim any of these
elements separately as new. The organizational contribution is to keep
the prepared state, a general dynamical invariant, and the
instantaneous energy basis logically distinct, and to express the
basis-specified occupation as their projector overlap.

The specific analytical results obtained from this organization are
the exact de Sitter scaling law
$N_k^{\mathrm{inst}}(\eta)=F_\nu(-k\eta)$, the mass-controlled
nonadiabatic scale $p_{\mathrm{phys,peak}}=m/\sqrt{2}$, and the
small-mass boundary layer that resolves the noncommuting massless and
late-time limits. These results do not define a new de Sitter spectrum.
They identify structure in the complete finite-time problem that is
obscured when different comoving modes are treated as independent
numerical trajectories.

The Bloch representation is especially useful when the complete
trajectory is of interest. The final Bogoliubov coefficient gives the
asymptotic occupation efficiently, whereas the real state vector makes
alignment, lag, transient overshoot, and coherent rotation directly
visible. The three independently integrated representations in
Sec.~\ref{sec:numerical_validation} also provide a finite-time
implementation check rather than only an asymptotic comparison.

The Lewis--Riesenfeld invariant remains useful for organizing exact
solutions, but a general invariant is not itself the prepared physical
state. In the aligned construction used here, the selected invariant
vector and the state Bloch vector satisfy the same initial-value problem
and therefore coincide. An invariant initialized in another direction
remains mathematically valid but does not encode the prepared state.
The occupation is determined by the physical density matrix projected
onto the instantaneous negative-energy subspace, not by an arbitrary
invariant component.

\subsection{Interpretational scope}

A particle concept at finite time is not unique in a general
time-dependent spacetime. It requires a specified instantaneous basis
or adiabatic prescription. In this work, the finite-time occupation is
defined relative to the instantaneous zeroth-order eigenbasis of the
mode Hamiltonian.

The transient overshoot and coherent ringdown in
Fig.~\ref{fig:universal_trajectories} should be understood within this
prescription. Their appearance in the Cartesian, angular, and
Bogoliubov calculations shows that they are not numerical artifacts.
Their detailed finite-time form, however, may change if a different
particle basis is chosen. The geometric formulation resolves the state
trajectory associated with a given prescription; it does not remove
the general ambiguity of finite-time particle number.

The asymptotic occupation has a stronger physical status. When
well-defined early- and late-time positive-frequency sectors exist,
different intermediate representations lead to the same final
Bogoliubov coefficient. In the de Sitter benchmark, the
Gibbons--Hawking plateau is therefore more robust than the detailed
trajectory approaching it.

A similar qualification applies to the entropy discussed above. The
complete two-mode particle--antiparticle state remains pure under
unitary evolution. The nonzero binary entropy is obtained only after
one member of the pair is traced out. It should therefore be
interpreted as particle--antiparticle reduced entanglement entropy, not
as the entropy of an isolated pure two-level state. This bipartite
interpretation is essential in both the finite-time and asymptotic
limits.

\subsection{Computational scope and limitations}

The Cartesian Bloch equation provides the most robust implementation
of the geometric dynamics. It is globally regular, evolves only real variables, and allows norm drift to be monitored directly. The angular representation preserves unit norm algebraically, but its coordinate equations become ill-conditioned near the poles of the Bloch sphere. This is precisely the region approached by a deep-subhorizon vacuum.
For this reason, the angular equations were used as an independent cross-check rather than as the primary numerical method.

The Bogoliubov formulation remains efficient and natural when the main
objective is the final occupation number. The geometric formulation is most useful when one also wants the
finite-time state trajectory, relative orientation, or nonadiabatic region. Neither representation is universally superior. Their usefulness depends on the observable and on whether intermediate-time information is required.

The convergence analysis also shows that a formally early initial time
is not always numerically sufficient. Heavy modes are particularly
sensitive to the finite starting point. Reliable calculations must
therefore report the initial and final integration boundaries,
tolerances, and convergence tests, rather than relying only on norm
preservation. Agreement between independent representations provides a
stronger diagnostic than the normalization condition alone.

\subsection{Power-law FLRW illustration}
\label{subsec:power-law-example}

To demonstrate explicitly that the construction is not restricted to
de Sitter space, consider the expanding power-law background
\begin{equation}
    a(\eta)
    =
    a_\star
    \left(\frac{\eta}{\eta_\star}\right)^q,
    \qquad
    \eta>0,
    \qquad
    q>0.
    \label{eq:power-law-scale-factor}
\end{equation}
This includes radiation- and matter-dominated expansion for $q=1$ and
$q=2$, respectively. Define the mode-dependent time at which
$ma(\eta_k)=k$ by
\begin{equation}
    \eta_k
    =
    \eta_\star
    \left(\frac{k}{ma_\star}\right)^{1/q},
    \qquad
    u=\frac{\eta}{\eta_k},
    \qquad
    \kappa_k=k\eta_k.
    \label{eq:power-law-variables}
\end{equation}
The Hamiltonian and exact Bloch equation then reduce to
\begin{equation}
    \hat H_k
    =
    k\left(u^q\sigma_1+\sigma_3\right),
    \qquad
    \frac{d\mathbf n_k}{du}
    =
    2\kappa_k
    \left(u^q,0,1\right)
    \times\mathbf n_k.
    \label{eq:power-law-bloch}
\end{equation}
The corresponding instantaneous occupation is obtained without any
further construction:
\begin{equation}
    N_k^{\mathrm{inst}}(u)
    =
    \frac{1}{2}
    \left[
        1-
        \frac{(u^q,0,1)}{\sqrt{1+u^{2q}}}
        \cdot\mathbf n_k(u)
    \right].
    \label{eq:power-law-occupation}
\end{equation}

This example also shows precisely which part of the de Sitter result is
special. In a power-law background the dynamics depends on the
additional mode parameter
$\kappa_k=k\eta_\star[k/(ma_\star)]^{1/q}$; hence the comoving
wavenumber does not collapse into a universal trajectory of $u$ alone.
The nonadiabaticity measure is
\begin{equation}
    \epsilon_{\mathrm{na}}^{(q)}(u)
    =
    \frac{q}{2\kappa_k}
    \frac{u^{q-1}}{\left(1+u^{2q}\right)^{3/2}}.
    \label{eq:power-law-nonadiabaticity}
\end{equation}
For $q>1$, its maximum occurs at
\begin{equation}
    u_{\mathrm{peak}}
    =
    \left(
        \frac{q-1}{2q+1}
    \right)^{1/(2q)},
    \label{eq:power-law-peak}
\end{equation}
while for $q=1$ the maximum lies at the initial boundary $u=0$.
For $0<q<1$, the measure instead grows toward the initial singular
boundary. For example, matter-dominated expansion gives
$u_{\mathrm{peak}}=5^{-1/4}$. Thus the same geometric equations apply,
but both the scaling structure and the location of the strongest basis
rotation differ from the de Sitter result.

\subsection{General scope beyond the benchmark}

The two-level $\mathrm{SU}(2)$ structure is not restricted to de
Sitter spacetime. After the standard mode reduction, a Dirac field in a
spatially flat FLRW background retains a time-dependent two-level
Hamiltonian. The physical Bloch vector and the geometric projection
onto an instantaneous particle basis can therefore be constructed for
a general scale factor.

The exact scaling found in the present benchmark is more restrictive.
The reduction to a universal function of $-k\eta$, the
momentum-independent late-time plateau, and the small-mass boundary
layer follow from the special scaling symmetry of exact de Sitter
space. These results should not be transferred unchanged to a general
FLRW background.

In less symmetric backgrounds, the geometric method may still be
useful even when no closed-form spectrum exists. Possible applications
include quasi-de Sitter expansion, reheating and fermionic preheating,
bouncing cosmologies, time-dependent effective masses, and fermion
production in external electric fields. In these cases, the driving
direction changes according to the specific background, while the
compact state-space representation remains available.

The de Sitter calculation should therefore be viewed as a controlled
benchmark. It demonstrates that the geometric formulation reproduces
the established asymptotic occupation, retains the full finite-time
trajectory, identifies the relevant non-adiabatic scale, and exposes a
singular small-mass boundary layer. At the same time, it makes clear
which statements are general consequences of fermionic
$\mathrm{SU}(2)$ dynamics and which rely on the special symmetry of de
Sitter spacetime.


\section{Conclusion}
\label{sec:conclusion}

We have developed a geometric formulation of fermionic particle creation in time-dependent curved spacetime based on the exact $\mathrm{SU}(2)$ structure of each Dirac momentum mode. The physical mode state is represented by a Bloch vector, while the instantaneous Hamiltonian defines a moving reference direction. Particle creation is
then expressed as the relative orientation between these two vectors. This representation is exactly equivalent to the standard Bogoliubov description, but it retains the complete finite-time state trajectory in a compact real geometry.

Within this framework, the fermionic occupation bound becomes
geometrically manifest. Its microscopic origin remains the canonical
anticommutation relations, while the compact Bloch sphere provides a
direct realization of the allowed occupation range. The formulation
also distinguishes clearly between the physical state vector and a
general Lewis--Riesenfeld invariant direction. In the aligned
construction used here, they coincide because they satisfy the same
precession equation with the same initial data; an invariant with
different initial data does not represent the prepared state.

The de Sitter application revealed several useful analytical
structures. The mode dynamics reduce exactly to a universal function of the dimensionless physical momentum and the mass-to-Hubble ratio, so different comoving momenta correspond to rescaled versions of the
same trajectory. The non-adiabatic response is controlled by a mass-dependent physical scale rather than by horizon crossing alone.
The analysis also exposes a singular small-mass boundary layer, which
resolves the apparent tension between exact conformal suppression at
zero mass and the finite late-time occupation obtained when the
positive-mass limit is taken after the asymptotic-time limit.

The numerical calculation confirms these results using independently
integrated Cartesian Bloch, angular Bloch, and Bogoliubov
representations. They agree throughout the finite-time evolution,
while explicit initial-boundary, final-boundary, tolerance, and
step-size studies establish numerical convergence. The transient
overshoot and coherent ringdown are not integration artifacts,
although their detailed form remains dependent on the instantaneous
particle basis.

The main value of the formulation is not a replacement of Bogoliubov
theory or a new de Sitter spectrum. It is a unified geometric
organization of fermionic mode evolution that makes normalization,
state-space compactness, finite-time motion, and particle projection
transparent within one structure. The power-law FLRW example
demonstrates explicitly that the same construction applies beyond de
Sitter space, while also showing that the universal de Sitter momentum
collapse does not persist in a general background. The formulation can
likewise be extended to time-dependent fermion masses, reheating and
bouncing cosmologies, and external-field particle production. In
backgrounds without closed-form Bogoliubov coefficients, the geometric
trajectory may provide a useful diagnostic of nonadiabatic evolution
and basis-specified particle--antiparticle entanglement.


\section*{Acknowledgements}

This work was supported by the Department of Science and Technology, Nampong Pattanasuksa Ratchamangklapisek School.


\section*{Data Availability Statement}

The Jupyter notebook, Python source code, and numerical tables used to
generate the figures and convergence results are provided as
Supplementary Material with this article. The calculations use only the equations and initial conditions stated in the manuscript and require no external datasets.


\appendix

\section{Exact de Sitter Spinor and the Late-Time Thermal Plateau}
\label{app:whittaker}

In this appendix, we derive the late-time fermionic occupation directly
from the complete two-component spinor. This is necessary because the
two asymptotic branches of a single decoupled scalar component cannot
be identified separately with the fermionic Bogoliubov coefficients.
The coefficients must instead be obtained by reconstructing the full
spinor and projecting it onto a normalized late-time basis.

\subsection{Two-component Dirac system}

After the standard conformal rescaling of the Dirac field and the
Fourier convention
\begin{equation}
    \chi(\eta,\mathbf{x})
    =
    \int\frac{d^3k}{(2\pi)^{3/2}}\,
    e^{i\mathbf{k}\cdot\mathbf{x}}\,
    \chi_{\mathbf{k}}(\eta),
\end{equation}
each momentum and helicity sector can be reduced to the two-component
system
\begin{equation}
    i\frac{d}{d\eta}
    \begin{pmatrix}
        q_+ \\
        q_-
    \end{pmatrix}
    =
    \begin{pmatrix}
        M(\eta) & k \\
        k & -M(\eta)
    \end{pmatrix}
    \begin{pmatrix}
        q_+ \\
        q_-
    \end{pmatrix},
    \qquad
    M(\eta)=m a(\eta).
    \label{eq:app_first_order_spinor}
\end{equation}

This representation is related to the Hamiltonian convention in
Eq.~\eqref{eq:hamiltonian_su2} by the constant unitary transformation
\begin{equation}
    U
    =
    \frac{1}{\sqrt{2}}
    \begin{pmatrix}
        1 & 1\\
        1 & -1
    \end{pmatrix},
    \qquad
    U\left(k\sigma_3+M\sigma_1\right)U^\dagger
    =
    M\sigma_3+k\sigma_1.
    \label{eq:app_constant_unitary}
\end{equation}
Because $U$ is time independent, it changes neither the
positive-frequency convention nor the occupation obtained from the
transformed energy projectors. The opposite helicity sector differs
only by a constant basis convention and gives the same occupation.

For the expanding de Sitter patch,
\begin{equation}
    a(\eta)
    =
    -\frac{1}{H\eta},
    \qquad
    \eta<0,
\end{equation}
we define
\begin{equation}
    \nu=\frac{m}{H},
    \qquad
    M(\eta)=-\frac{\nu}{\eta}.
    \label{eq:app_mass_ratio}
\end{equation}
Equation~\eqref{eq:app_first_order_spinor} then becomes
\begin{align}
    iq_+'
    &=
    -\frac{\nu}{\eta}q_+
    +kq_-,
    \label{eq:app_qplus_first}
    \\
    iq_-'
    &=
    kq_+
    +\frac{\nu}{\eta}q_-.
    \label{eq:app_qminus_first}
\end{align}

Decoupling the two components gives
\begin{align}
    q_+''
    +
    \left[
        k^2+
        \frac{\nu^2+i\nu}{\eta^2}
    \right]q_+
    &=
    0,
    \label{eq:app_qplus_second}
    \\
    q_-''
    +
    \left[
        k^2+
        \frac{\nu^2-i\nu}{\eta^2}
    \right]q_-
    &=
    0.
    \label{eq:app_qminus_second}
\end{align}
The opposite signs of the imaginary terms are essential. They ensure
that the two scalar equations remain the components of one normalized
first-order Dirac spinor.

\subsection{Whittaker representation and positive frequency}

We choose the Whittaker variable
\begin{equation}
    z=2ik\eta.
    \label{eq:app_whittaker_variable}
\end{equation}
Since $\eta<0$ and $k>0$, this variable lies on the ray
$\arg z=-\pi/2$. Equations~\eqref{eq:app_qplus_second} and
\eqref{eq:app_qminus_second} take the standard Whittaker form with Whittaker parameter $\kappa=0$ \cite{Olver2010},
\begin{equation}
    \frac{d^2q_\pm}{dz^2}
    +
    \left[
        -\frac{1}{4}
        +
        \frac{\frac{1}{4}-\lambda_\pm^2}{z^2}
    \right]q_\pm
    =
    0,
    \label{eq:app_whittaker_equation}
\end{equation}
where
\begin{equation}
    \lambda_+
    =
    \frac{1}{2}-i\nu,
    \qquad
    \lambda_-
    =
    \frac{1}{2}+i\nu.
    \label{eq:app_whittaker_indices}
\end{equation}
Indeed,
\begin{equation}
    \frac{1}{4}-\lambda_+^2
    =
    \nu^2+i\nu,
    \qquad
    \frac{1}{4}-\lambda_-^2
    =
    \nu^2-i\nu,
\end{equation}
in agreement with Eqs.~\eqref{eq:app_qplus_second} and
\eqref{eq:app_qminus_second}.

The positive-frequency convention used throughout this work is
\begin{equation}
    q_k^{(+)}(\eta)
    \sim
    e^{-ik\eta},
    \qquad
    \eta\rightarrow-\infty.
    \label{eq:app_positive_frequency}
\end{equation}
This convention fixes the sign in
Eq.~\eqref{eq:app_whittaker_variable}, since
\begin{equation}
    W_{0,\lambda}(z)
    \sim
    e^{-z/2}
    =
    e^{-ik\eta},
    \qquad
    |z|\rightarrow\infty.
\end{equation}

A normalized early-time positive-frequency spinor is therefore
\begin{equation}
    Q_k^{\mathrm{in}}(\eta)
    =
    \frac{1}{\sqrt{2}}
    \begin{pmatrix}
        W_{0,\frac{1}{2}-i\nu}(2ik\eta)
        \\
        W_{0,\frac{1}{2}+i\nu}(2ik\eta)
    \end{pmatrix}.
    \label{eq:app_exact_in_spinor}
\end{equation}
The two components are not independent. The Whittaker identity
\begin{equation}
    \left(
        -2\frac{d}{dz}
        +
        \frac{2i\nu}{z}
    \right)
    W_{0,\frac{1}{2}-i\nu}(z)
    =
    W_{0,\frac{1}{2}+i\nu}(z)
    \label{eq:app_whittaker_identity}
\end{equation}
shows that Eq.~\eqref{eq:app_exact_in_spinor} satisfies the original
first-order system.

At early times,
\begin{equation}
    Q_k^{\mathrm{in}}(\eta)
    \longrightarrow
    \frac{e^{-ik\eta}}{\sqrt{2}}
    \begin{pmatrix}
        1\\
        1
    \end{pmatrix}.
    \label{eq:app_early_spinor}
\end{equation}
This is the normalized positive-energy eigenstate of the asymptotic
Hamiltonian $k\sigma_1$. The normalization is preserved at all times
by the Hermitian first-order evolution.

\subsection{Late-time spinor and Bogoliubov projection}

For $\nu>0$ and small $z$ on the branch fixed above, the Whittaker
function has the expansion
\begin{equation}
    W_{0,\lambda}(z)
    \sim
    \frac{\Gamma(-2\lambda)}
         {\Gamma\left(\frac{1}{2}-\lambda\right)}
    z^{\frac{1}{2}+\lambda}
    +
    \frac{\Gamma(2\lambda)}
         {\Gamma\left(\frac{1}{2}+\lambda\right)}
    z^{\frac{1}{2}-\lambda}.
    \label{eq:app_small_z_whittaker}
\end{equation}
Using Eq.~\eqref{eq:app_whittaker_indices}, the two spinor components
behave as
\begin{align}
    q_+^{\mathrm{in}}
    &\sim
    \frac{1}{\sqrt{2}}\,
    \frac{\Gamma(1-2i\nu)}
         {\Gamma(1-i\nu)}
    z^{i\nu}
    +
    O\left(z^{1-i\nu}\right),
    \label{eq:app_late_qplus}
    \\
    q_-^{\mathrm{in}}
    &\sim
    \frac{1}{\sqrt{2}}\,
    \frac{\Gamma(1+2i\nu)}
         {\Gamma(1+i\nu)}
    z^{-i\nu}
    +
    O\left(z^{1+i\nu}\right).
    \label{eq:app_late_qminus}
\end{align}
The omitted terms vanish as $\eta\rightarrow0^-$. The two surviving
terms belong to different components of the full spinor and can
therefore be projected onto the positive- and negative-energy
late-time states.

Define
\begin{equation}
    r=-2k\eta>0.
\end{equation}
Along the de Sitter time contour,
\begin{equation}
    z=-ir.
\end{equation}
Using the principal branch of the logarithm gives
\begin{equation}
    z^{i\nu}
    =
    e^{\pi\nu/2}r^{i\nu},
    \qquad
    z^{-i\nu}
    =
    e^{-\pi\nu/2}r^{-i\nu}.
    \label{eq:app_branch_factors}
\end{equation}

At late times, $M(\eta)\gg k$, and the normalized instantaneous
eigenvectors approach the out-basis
\begin{equation}
    U_k^{\mathrm{out}}(\eta)
    \sim
    \begin{pmatrix}
        r^{i\nu}\\
        0
    \end{pmatrix},
    \qquad
    V_k^{\mathrm{out}}(\eta)
    \sim
    \begin{pmatrix}
        0\\
        r^{-i\nu}
    \end{pmatrix}.
    \label{eq:app_out_basis}
\end{equation}
The corrections to these basis vectors are of relative order $k/M$ and
vanish as $\eta\rightarrow0^-$. The first vector has positive
instantaneous energy, while the second has negative instantaneous
energy.

The in-mode decomposes asymptotically as
\begin{equation}
    Q_k^{\mathrm{in}}
    \longrightarrow
    \alpha_k U_k^{\mathrm{out}}
    +
    \beta_k V_k^{\mathrm{out}}.
    \label{eq:app_bogoliubov_projection}
\end{equation}
Equivalently, the asymptotic projection coefficients are
\begin{equation}
    \alpha_k
    =
    \lim_{r\rightarrow0^+}
    r^{-i\nu}q_+^{\mathrm{in}},
    \qquad
    \beta_k
    =
    \lim_{r\rightarrow0^+}
    r^{i\nu}q_-^{\mathrm{in}}.
    \label{eq:app_projection_limits}
\end{equation}
Up to irrelevant phase conventions for the out-basis,
Eqs.~\eqref{eq:app_late_qplus}--\eqref{eq:app_branch_factors} give
\begin{align}
    \alpha_k
    &=
    \frac{e^{\pi\nu/2}}{\sqrt{2}}\,
    \frac{\Gamma(1-2i\nu)}
         {\Gamma(1-i\nu)},
    \label{eq:app_alpha_exact}
    \\
    \beta_k
    &=
    \frac{e^{-\pi\nu/2}}{\sqrt{2}}\,
    \frac{\Gamma(1+2i\nu)}
         {\Gamma(1+i\nu)}.
    \label{eq:app_beta_exact}
\end{align}

For real $\nu$, the two Gamma-function ratios have equal magnitude.
Using
\begin{equation}
    \left|\Gamma(1+iy)\right|^2
    =
    \frac{\pi y}{\sinh(\pi y)},
    \label{eq:app_gamma_identity}
\end{equation}
we obtain
\begin{equation}
    \left|
        \frac{\Gamma(1-2i\nu)}
             {\Gamma(1-i\nu)}
    \right|^2
    =
    \left|
        \frac{\Gamma(1+2i\nu)}
             {\Gamma(1+i\nu)}
    \right|^2
    =
    \frac{1}{\cosh(\pi\nu)}.
    \label{eq:app_gamma_ratio}
\end{equation}
Consequently,
\begin{align}
    |\alpha_k|^2
    &=
    \frac{e^{\pi\nu}}
         {2\cosh(\pi\nu)}
    =
    \frac{1}{1+e^{-2\pi\nu}},
    \label{eq:app_alpha_squared}
    \\
    |\beta_k|^2
    &=
    \frac{e^{-\pi\nu}}
         {2\cosh(\pi\nu)}
    =
    \frac{1}{e^{2\pi\nu}+1}.
    \label{eq:app_beta_squared}
\end{align}
The fermionic normalization follows directly:
\begin{equation}
    |\alpha_k|^2+|\beta_k|^2=1.
    \label{eq:app_bogoliubov_normalization}
\end{equation}
Equivalently,
\begin{equation}
    \left|
        \frac{\beta_k}{\alpha_k}
    \right|^2
    =
    e^{-2\pi\nu}.
    \label{eq:app_bogoliubov_ratio}
\end{equation}

The late-time occupation is therefore
\begin{equation}
    N_k^{\mathrm{out}}
    =
    |\beta_k|^2
    =
    \frac{1}
    {\exp\left(2\pi m/H\right)+1}.
    \label{eq:app_gh_plateau}
\end{equation}
This is the Fermi--Dirac factor associated with the de Sitter
temperature
\begin{equation}
    T_{\mathrm{dS}}
    =
    \frac{H}{2\pi}.
\end{equation}
The result is independent of the comoving momentum because every fixed
mode is redshifted to vanishing physical momentum in the late-time
limit.

The derivation assumes $\nu>0$ and takes the late-time limit before the
massless limit. The strictly massless theory remains conformally
invariant and has zero particle production. The difference between
these two limiting procedures is resolved by the boundary-layer
behavior discussed in the main text.

\bibliographystyle{unsrtnat}
\bibliography{references}

@article{Parker1968,
  author  = {Parker, Leonard},
  title   = {Particle Creation in Expanding Universes},
  journal = {Physical Review Letters},
  volume  = {21},
  number  = {8},
  pages   = {562--564},
  year    = {1968},
  doi     = {10.1103/PhysRevLett.21.562}
}

@article{Parker1969,
  author  = {Parker, Leonard},
  title   = {Quantized Fields and Particle Creation in Expanding Universes. I},
  journal = {Physical Review},
  volume  = {183},
  number  = {5},
  pages   = {1057--1068},
  year    = {1969},
  doi     = {10.1103/PhysRev.183.1057}
}

@article{Fulling1973,
  author  = {Fulling, Stephen A.},
  title   = {Nonuniqueness of Canonical Field Quantization in Riemannian Space-Time},
  journal = {Physical Review D},
  volume  = {7},
  number  = {10},
  pages   = {2850--2862},
  year    = {1973},
  doi     = {10.1103/PhysRevD.7.2850}
}

@book{BirrellDavies,
  author    = {Birrell, N. D. and Davies, P. C. W.},
  title     = {Quantum Fields in Curved Space},
  publisher = {Cambridge University Press},
  address   = {Cambridge},
  year      = {1982},
  doi       = {10.1017/CBO9780511622632}
}

@book{ParkerToms,
  author    = {Parker, Leonard and Toms, David},
  title     = {Quantum Field Theory in Curved Spacetime: Quantized Fields and Gravity},
  publisher = {Cambridge University Press},
  address   = {Cambridge},
  year      = {2009},
  doi       = {10.1017/CBO9780511813924}
}

@article{Ford2021,
  author  = {Ford, L. H.},
  title   = {Cosmological Particle Production: A Review},
  journal = {Reports on Progress in Physics},
  volume  = {84},
  number  = {11},
  pages   = {116901},
  year    = {2021},
  doi     = {10.1088/1361-6633/ac1b23}
}

@article{GibbonsHawking1977,
  author  = {Gibbons, G. W. and Hawking, S. W.},
  title   = {Cosmological Event Horizons, Thermodynamics, and Particle Creation},
  journal = {Physical Review D},
  volume  = {15},
  number  = {10},
  pages   = {2738--2751},
  year    = {1977},
  doi     = {10.1103/PhysRevD.15.2738}
}

@article{Greene1999,
  author  = {Greene, Patrick B. and Kofman, Lev},
  title   = {Preheating of Fermions},
  journal = {Physics Letters B},
  volume  = {448},
  number  = {1--2},
  pages   = {6--12},
  year    = {1999},
  doi     = {10.1016/S0370-2693(99)00020-9}
}

@article{Adshead2015,
  author        = {Adshead, Peter and Sfakianakis, Evangelos I.},
  title         = {Fermion Production during and after Axion Inflation},
  journal       = {Journal of Cosmology and Astroparticle Physics},
  volume        = {2015},
  number        = {11},
  pages         = {021},
  year          = {2015},
  doi           = {10.1088/1475-7516/2015/11/021},
  eprint        = {1508.00891},
  archivePrefix = {arXiv},
  primaryClass  = {hep-ph}
}

@article{Domcke2018,
  author        = {Domcke, Valerie and Mukaida, Kyohei},
  title         = {Gauge Field and Fermion Production during Axion Inflation},
  journal       = {Journal of Cosmology and Astroparticle Physics},
  volume        = {2018},
  number        = {11},
  pages         = {020},
  year          = {2018},
  doi           = {10.1088/1475-7516/2018/11/020},
  eprint        = {1806.08769},
  archivePrefix = {arXiv},
  primaryClass  = {hep-ph}
}

@article{Cotaescu2002,
  author        = {Cot\u{a}escu, Ion I.},
  title         = {Polarized Dirac Fermions in de Sitter Spacetime},
  journal       = {Physical Review D},
  volume        = {65},
  number        = {8},
  pages         = {084008},
  year          = {2002},
  doi           = {10.1103/PhysRevD.65.084008},
  eprint        = {hep-th/0109199},
  archivePrefix = {arXiv}
}

@article{Stahl2016,
  author        = {Stahl, Cl{\'e}ment and Strobel, Eckhard and Xue, She-Sheng},
  title         = {Fermionic Current and Schwinger Effect in de Sitter Spacetime},
  journal       = {Physical Review D},
  volume        = {93},
  number        = {2},
  pages         = {025004},
  year          = {2016},
  doi           = {10.1103/PhysRevD.93.025004},
  eprint        = {1507.01686},
  archivePrefix = {arXiv},
  primaryClass  = {gr-qc}
}

@article{ParkerFulling1974,
  author  = {Parker, Leonard and Fulling, Stephen A.},
  title   = {Adiabatic Regularization of the Energy-Momentum Tensor of a Quantized Field in Homogeneous Spaces},
  journal = {Physical Review D},
  volume  = {9},
  number  = {2},
  pages   = {341--354},
  year    = {1974},
  doi     = {10.1103/PhysRevD.9.341}
}

@article{Landete2014,
  author        = {Landete, A. and Navarro-Salas, J. and Torrent{\'i}, F.},
  title         = {Adiabatic Regularization for Spin-1/2 Fields},
  journal       = {Physical Review D},
  volume        = {89},
  number        = {4},
  pages         = {044030},
  year          = {2014},
  doi           = {10.1103/PhysRevD.89.044030},
  eprint        = {1311.4958},
  archivePrefix = {arXiv},
  primaryClass  = {gr-qc}
}

@article{DabrowskiDunne2016,
  author        = {Dabrowski, Robert and Dunne, Gerald V.},
  title         = {Time Dependence of Adiabatic Particle Number},
  journal       = {Physical Review D},
  volume        = {94},
  number        = {6},
  pages         = {065005},
  year          = {2016},
  doi           = {10.1103/PhysRevD.94.065005},
  eprint        = {1606.00902},
  archivePrefix = {arXiv},
  primaryClass  = {hep-th}
}

@article{Ferreiro2022,
  author        = {Ferreiro, Antonio and Navarro-Salas, Jose},
  title         = {Pair creation in electric fields, anomalies, and renormalization of the electric current},
  journal       = {Physical Review D},
  volume        = {97},
  number        = {12},
  pages         = {125012},
  year          = {2018},
  doi           = {10.1103/PhysRevD.97.125012}
}

@article{Feynman1957,
  author  = {Feynman, Richard P. and Vernon, Jr., Frank L. and Hellwarth, Robert W.},
  title   = {Geometrical Representation of the Schr\"odinger Equation for Solving Maser Problems},
  journal = {Journal of Applied Physics},
  volume  = {28},
  number  = {1},
  pages   = {49--52},
  year    = {1957},
  doi     = {10.1063/1.1722572}
}

@article{BialynickiBirula1991,
  author  = {Bia\l{}ynicki-Birula, Iwo and G\'ornicki, Pawe\l{} and Rafelski, Johann},
  title   = {Phase-Space Structure of the Dirac Vacuum},
  journal = {Physical Review D},
  volume  = {44},
  number  = {6},
  pages   = {1825--1835},
  year    = {1991},
  doi     = {10.1103/PhysRevD.44.1825}
}

@article{Lewis1969,
  author  = {Lewis, H. R. and Riesenfeld, W. B.},
  title   = {An Exact Quantum Theory of the Time-Dependent Harmonic Oscillator and of a Charged Particle in a Time-Dependent Electromagnetic Field},
  journal = {Journal of Mathematical Physics},
  volume  = {10},
  number  = {8},
  pages   = {1458--1473},
  year    = {1969},
  doi     = {10.1063/1.1664991}
}

@article{Seger1995,
  author        = {Seger, J. E. and Balantekin, A. B.},
  title         = {Fermion Pair Production from an Electric Field Varying in Two Dimensions},
  journal       = {Journal of Mathematical Physics},
  volume        = {37},
  number        = {1},
  pages         = {219--226},
  year          = {1996},
  doi           = {10.1063/1.531385},
  eprint        = {nucl-th/9506022},
  archivePrefix = {arXiv}
}

@article{Kim1999,
  author        = {Kim, Sang Pyo and Santana, A. E. and Khanna, F. C.},
  title         = {Generalized Invariants and Quantum Evolution of Open Fermionic Systems},
  journal       = {Physics Letters A},
  volume        = {272},
  number        = {1--2},
  pages         = {46--52},
  year          = {2000},
  doi           = {10.1016/S0375-9601(00)00406-0},
  eprint        = {hep-th/9910189},
  archivePrefix = {arXiv}
}

@article{Khantoul2015,
  author        = {Khantoul, Boubakeur and Fring, Andreas},
  title         = {Time-Dependent Massless Dirac Fermions in Graphene},
  journal       = {Physics Letters A},
  volume        = {379},
  number        = {42},
  pages         = {2704--2706},
  year          = {2015},
  doi           = {10.1016/j.physleta.2015.08.011},
  eprint        = {1505.02087},
  archivePrefix = {arXiv},
  primaryClass  = {quant-ph}
}

@article{Min2019,
  author        = {Min, Ui and Son, Minho and Suh, Han Gyeol},
  title         = {Group Theoretic Approach to Fermion Production},
  journal       = {Journal of High Energy Physics},
  volume        = {2019},
  number        = {3},
  pages         = {072},
  year          = {2019},
  doi           = {10.1007/JHEP03(2019)072},
  eprint        = {1808.00939},
  archivePrefix = {arXiv},
  primaryClass  = {hep-ph}
}

@article{Gong2025,
  author        = {Gong, Jia-Qi and Yang, Ji-Chong},
  title         = {Digit Quantum Simulation of a Fermion Field in an Expanding Universe},
  journal       = {Physical Review D},
  volume        = {112},
  number        = {9},
  pages         = {096020},
  year          = {2025},
  doi           = {10.1103/btfs-m58w},
  eprint        = {2502.14021},
  archivePrefix = {arXiv},
  primaryClass  = {quant-ph}
}

@article{Alkhateeb2026,
  author        = {Alkhateeb, Mohammed and Edwards, James P. and Caudano, Yves},
  title         = {Computational Quantum Field Theory for Fermion Pair Creation in Two-Dimensional Curved Spacetimes},
  journal       = {Physical Review D},
  volume        = {113},
  number        = {12},
  pages         = {125002},
  year          = {2026},
  doi           = {10.1103/gb8d-kqc5},
  eprint        = {2602.07330},
  archivePrefix = {arXiv},
  primaryClass  = {hep-th}
}

@book{Olver2010,
  editor    = {Olver, Frank W. J. and Lozier, Daniel W. and Boisvert, Ronald F. and Clark, Charles W.},
  title     = {NIST Handbook of Mathematical Functions},
  publisher = {Cambridge University Press},
  address   = {Cambridge},
  year      = {2010}
}

@book{Hairer1993,
  author    = {Hairer, Ernst and N{\o}rsett, Syvert P. and Wanner, Gerhard},
  title     = {Solving Ordinary Differential Equations I: Nonstiff Problems},
  edition   = {2},
  publisher = {Springer},
  address   = {Berlin},
  year      = {1993},
  doi       = {10.1007/978-3-540-78862-1}
}

@article{GueryOdelin2019,
  author  = {Gu{\'e}ry-Odelin, David and Ruschhaupt, Andreas and
             Kiely, Anthony and Torrontegui, Erik and
             Mart{\'i}nez-Garaot, Sof{\'i}a and Muga, Juan Gonzalo},
  title   = {Shortcuts to Adiabaticity: Concepts, Methods, and Applications},
  journal = {Reviews of Modern Physics},
  volume  = {91},
  number  = {4},
  pages   = {045001},
  year    = {2019},
  doi     = {10.1103/RevModPhys.91.045001}
}

@article{Fuentes2010,
  author  = {Fuentes, Ivette and Mann, Robert B. and
             Mart{\'i}n-Mart{\'i}nez, Eduardo and Moradi, Shahpoor},
  title   = {Entanglement of Dirac Fields in an Expanding Spacetime},
  journal = {Physical Review D},
  volume  = {82},
  number  = {4},
  pages   = {045030},
  year    = {2010},
  doi     = {10.1103/PhysRevD.82.045030}
}

@article{Kanno2017,
  author        = {Kanno, Sugumi and Sasaki, Misao and Tanaka, Takahiro},
  title         = {Vacuum State of the Dirac Field in de Sitter Space
                   and Entanglement Entropy},
  journal       = {Journal of High Energy Physics},
  volume        = {2017},
  number        = {3},
  pages         = {068},
  year          = {2017},
  doi           = {10.1007/JHEP03(2017)068},
}

\end{document}